\documentclass[superscriptaddress,secnumarabic,
 amssymb,amsmath,nobibnotes,aps,prd,showkeys,showpacs]{revtex4}

\usepackage{graphicx,epsfig,amssymb,amsmath,amstext}

\usepackage{graphicx}
\usepackage{bm}%
\newcommand{\be}{\begin{equation}}
\newcommand{\ee}{\end{equation}}
\newcommand{\bea}{\begin{eqnarray}}

\newcommand{\eea}{\end{eqnarray}}

\begin{document}

\title{An observable signature of a background deviating from Kerr}

\author{Georgios Lukes-Gerakopoulos}
\affiliation{Academy of Athens, Research Center for Astronomy,
Soranou Efesiou 4, GR-11527, Athens, GREECE}
\affiliation{Section of Astrophysics, Astronomy, and Mechanics,
Department of Physics, University of Athens, Panepistimiopolis Zografos GR15783,
Athens, Greece}

\author{Theocharis A. Apostolatos}
\affiliation{Section of Astrophysics, Astronomy, and Mechanics,
Department of Physics, University of Athens, Panepistimiopolis Zografos GR15783,
Athens, Greece}

\author{George Contopoulos}
\affiliation{Academy of Athens, Research Center for Astronomy,
 Soranou Efesiou 4, GR-11527, Athens, GREECE}

\begin{abstract}
By detecting gravitational wave signals from extreme mass ratio inspiraling
sources (EMRIs) we will be given the opportunity to check our theoretical
expectations regarding the nature of supermassive bodies that inhabit the
central regions of galaxies. We have explored some qualitatively new features
that a perturbed Kerr metric induces in its geodesic orbits. Since a generic
perturbed Kerr metric does not possess all the special symmetries of a Kerr
metric, the geodesic equations in the former case are described by a slightly
nonintegrable Hamiltonian system. According to the Poincar\'{e}-Birkhoff theorem
this causes the appearance of the so-called Birkhoff chains of islands on the
corresponding surfaces of section in between the anticipated KAM curves of the
integrable Kerr case, whenever the intrinsic frequencies of the system are at
resonance. The chains of islands are characterized by finite width, i.e. there
is a finite range of initial conditions that correspond to a particular
resonance and consequently to a constant rational ratio of intrinsic frequencies.
Thus while the EMRI changes adiabatically by radiating energy and angular
momentum, by monitoring the frequencies of a signal we can look for a transient
pattern, in the form of a plateau, in the evolution of their ratio. We have
shown that such a plateau is anticipated to be apparent in a quite large
fraction of possible orbital characteristics if the central gravitating source
is not a Kerr black hole. Moreover the plateau in the ratio of frequencies is
expected to be more prominent at specific rational values that correspond to the
strongest resonances. This gives a possible observational detection of such
non-Kerr exotic objects.
\end{abstract}

\pacs{04.30.-w; 97.60.Lf; 05.45.-a}
\keywords{Gravitational waves, black holes, KAM theorem}
\maketitle

\section{Introduction}

The enterprize of searching for gravitational waves with the interferometric
devices that are already operating in various places around the globe is under
way. Unfortunately no positive outcome has come up yet \cite{overview}.
Gravitational waves visiting us from powerful cosmic sources are very weak upon
reaching our planet. The sensitivity of the present day detectors is not
adequate to distinguish undoubtedly the signal from noise; it is improving
continuously though. On the other hand null detection of gravitational waves
puts firm constraints on various physical parameters related with known sources
of gravitational waves (e.g.~rotating neutron stars \cite{pulsars}).

LISA, the future interferometric detector that is planned to be launched by NASA
and ESA during the forthcoming decade, will be much more sensitive in detecting
gravitational waves from much more massive and much more distant sources
\cite{LISA}. Among such sources are EMRIs (extreme-mass-ratio inspirals), that
is low mass compact objects (neutron stars or solar-mass black holes) trapped in
the gravitational field of supermassive compact objects which are probably
located at the centers of galaxies \cite{EMRI,EMRIastro,tobeaKerr}. According to
conventional astrophysical wisdom, such supermassive objects are Kerr black
holes. The gravitational waves emitted in a Kerr background have been
extensively studied by a large number of people, and useful techniques have been
invented to gain physical information about the characteristics of these sources
from the analysis of the corresponding waveforms
\cite{KerrEMRI,GlamBaba06,MockData}. Of course the extraction of valuable
astrophysical information from the gravitational waves will be hindered by
instrumental noise \cite{LISA} and confusion noise \cite{noise} (background
unresolved signals). Furthermore practical difficulties concerning the search in
the multi-parametric space of templates for EMRIs that will be used to detect
such signals render the accomplishment of such a goal even harder
\cite{multiparameter}.

EMRIs are ideal sources of gravitational waves through which one could explore
the background metric that the lighter object is tracing. In 1995 Ryan
\cite{Ryan95,Ryan97} showed that almost circular and almost equatorial orbits in
a generic stationary and axisymmetric background could reveal the lower order
mass moments and mass-current moments by monitoring the corresponding evolution
of gravitational waves that are emitted in the weak field region. Since all
multiple moments of a Kerr black hole are determined only by its mass and its
spin, one needs to know at least the quadrupole moment (besides the mass and the
spin) of the central object to decide if the object is a Kerr metric or not.
However, while the measurement of the quadrupole moment is in principle feasible
with LISA, the accuracy of such measurement is limited due to a number of
reasons: (i) The orbits of EMRIs are expected to be much more general than
simply circular and equatorial. (ii) The analysis is limitted to work in the
weak field region, where the gravitational wave emission is not as intense
as in the strong field region. (iii) The source should not be a very distant
one, so that the signal to noise ratio in the analysis of gravitational waves is
sufficiently high to be able to obtain the parameters of the source with good
accuracy.

An alternative approach for checking whether the background metric is that of a
Kerr black hole is to focus on a characteristic signature of EMRIs in a generic
non-Kerr background. This could be a yes/no measurement from which one could
decide if the field into which the small object orbits is that of a massive Kerr
black hole or not. Various studies have been performed towards this line of
thought. Collins and Hughes \cite{CollHugh04} built a perturbed Kerr metric and
explored the effect of these perturbations on physically observable quantities
related with geodesics in such a background. Later, Glampedakis and Babak
\cite{GlamBaba06} showed that a perturbed Kerr metric leads to a significant
mismatch between the corresponding waveforms, although there is an issue of
confusion between waveforms of the specific non-Kerr metric and a Kerr metric
with different orbital parameters. A similar approach assuming a more
astrophysically oriented non-Kerr metric was followed by Barausse et al
\cite{Rezz}. Recently, Gair et al \cite{Gair08} used a specific exact solution
of vacuum Einstein equations, namely the Manko-Novikov solution \cite{Manko92},
which could be turned into a pure Kerr metric by dialing a single parameter, and
studied the geodesics in such a metric. Their analysis showed that for a range
of metric parameters there are two ring-like regions of bounded orbits in this
space-time. In the outer one the orbits look regular, as if there exists an
isolating integral of motion, while in the inner region the orbits seem to be
ergodic. The two regions, though, merge for a range of orbital parameters. It
was suggested by Gair et al \cite{Gair08} that the transition of the orbits from
the regular to the ergodic region when the two aforementioned regions merge,
could have a clear observable effect on the frequency spectrum of the
corresponding gravitational waves, especially if this transition is repetitive
from the former to the latter region and vice versa. Another effect that was
thoroughly explored by Gair et al is the instability that may arise just before
the orbiting body arrives at its innermost circular orbit (ISCO). This effect
might have a strong influence on the gravitational waveform at the final stage
of the corresponding EMRI. However, these new orbital features vary
significantly with the orbital and metric parameters.

Thus it would be important to seek for a generic feature that clearly
distinguishes a non-Kerr metric from a Kerr one. Then it would be much easier to
perform a specialized observation to measure it. The fact that Kerr black holes
are so perfect and symmetric objects which are created naturally, renders all
other gravitational fields of isolated massive objects quite different. In
contrast to the Kerr metric, any other generic solution of Einstein field
equations that describes the neighborhood of an isolated massive object is
expected to be less symmetric. Even axisymmetric and stationary metrics, but
otherwise generic ones, are not expected to have something analogous to the
Carter Killing tensor field, as it happens with the Kerr metric \cite{Carter68}.
Therefore the existence of a Carter-like constant that makes the geodesic orbits
in a Kerr metric to be characterized by a completely integrable system of
equations \cite{Chandra}, most probably is lost in a non-Kerr axially symmetric
and stationary, asymptotically flat metric. Geodesic orbits in such a generic
metric is described by a nonintegrable Hamiltonian that deviates from the
corresponding integrable Kerr-type Hamiltonian. Furthermore, if such a non-Kerr
metric is not drastically different from a Kerr one, a generic axially symmetric,
and stationary metric, that could in principle be realized by natural processes,
could be considered as a perturbed Kerr metric.

The KAM theorem of Kolmogorov, Arnold and Moser \cite{KAM}, applies to
Hamiltonian systems that are slightly perturbed from integrable systems.
According to this general theorem most tori of the corresponding integrable
system, that is the toroidal hyper-surfaces of phase-space on which the phase
orbits of the system are lying, get slightly deformed in the perturbed system,
but they are not destroyed. This is exactly the case presented by Gair et al
\cite{Gair08}, who studied geodesic orbits in the so-called Manko-Novikov metric.

On the other hand according to the Poincar\'{e}-Birkhoff theorem, the resonant
tori (the tori characterized by a rational ratio of winding frequencies) of the
integrable system disintegrate when the system gets perturbed, and consequently
a chain of islands is formed on a surface of section, instead of an infinite
number of sets of periodic points which is a characteristic feature of an
integrable system. These islands are characterized, as in the resonant tori of
the integrable system, by a ratio of winding frequencies of the system that
equals a rational number. The appearance of such islands is a very distinct new
signature of a slightly nonintegrable system and it is qualitatively different
from the behavior of geodesics in a Kerr spacetime. They are always present
independently of the way the system deviates from the corresponding integrable
one. This new feature, if observable, could clearly distinguish any kind of
perturbed Kerr metric from a pure Kerr metric \cite{our}.

We start by showing that these Birkhoff islands are actually present
in the Manko-Novikov metric studied in \cite{Gair08}. However, such
islands were not found in \cite{Gair08}, because a very thorough
exploration of the phase-space is needed in order to reveal their
existence. The rotation number, a tool widely used in the study of
systems where chaos and order coexist, is employed in the search of
suitable initial orbital conditions that lead to the formation of
Birkhoff islands on a surface of section. The rotation number is an
index that is directly related to observable quantities (see below).

Next we argue that, since these islands can be identified by monitoring specific
quantities, namely the frequencies of the emitted gravitational waves, we can
tell if the metric in which the orbiting object is moving is a Kerr or a
non-Kerr one by not observing or observing, respectively, a fixed ratio of the
two polar frequencies for a period of time while the frequencies themselves are
evolving. An observed plateau in the evolution of the ratio of the corresponding
frequencies would be an unambiguous signature of a non-Kerr metric. Moreover
since the most pronounced resonances (thicker Birkhoff islands) correspond to
simple-integer-number ratios, like 2:3, or 1:2, the value of the ratio of
frequencies that is related to the most extended plateaus should be a
simple-integer-number ratio as well. Thus we can focus our signal analysis on
the time interval when the ratio of polar frequencies assumes that value and
investigate the existence or not of a plateau in the evolution of the ratio of
frequencies. Even a null result in such an analysis can be used to put firm
constraints in the likelihood that we have observed an EMRI in a pure Kerr
metric through gravitational waves.

Now from a practical point of view, we argue that a large fraction of EMRIs that
get trapped in the gravitational field of a supermassive compact object in a
generic eccentric and non-equatorial orbit will evolve in such a way that the
corresponding orbit will eventually cross a resonant torus (if the central
massive object is a Kerr black hole) or the corresponding Birkhoff chain of
islands (if the central object is described by a perturbed Kerr metric).
Respectively, the ratio of the frequencies will vary strictly monotonically, or
it will form a plateau. In the latter case the duration of the plateau will
depend (i) on the particular metric which is related to the characteristics of
the central massive object, (ii) on the orbital parameters of the low-mass
object, and (iii) on the ratio of the masses involved in the binary as well.
This plateau can in principle be identified for a source with a sufficiently
high signal-to-noise ratio.

The present paper is organized as follows. In Section \ref{sec:2} we repeat the
basic characteristics of the Manko-Novikov solution which was used and
thoroughly analyzed in \cite{Gair08} as an example of a non-Kerr metric which
could be transformed into a Kerr one by suitably adjusting a single parameter.
We are using the same metric to exhibit the new feature that we propose to use
as a yes/no test of a Kerr metric. In Sec.~\ref{sec:3} we start with a short
description of the Poincar\'{e}-Birkhoff theorem as a theoretical basis applied
in the perturbed Kerr case, and then we present a thorough study of the surfaces
of section of the specific non-Kerr metric; the Manko-Novikov one. We show how
we have determined the initial conditions so as to form a few Birkhoff chains of
islands, by implementing a numerical measurement of the rotation number for each
geodesic. Both regions of bound orbits that are present in this metric are
thoroughly explored through surfaces of section and the final picture is
discussed in every case; either when the two regions are separated or connected.
Apart from the Birkhoff chains of islands, a new characteristic found in our
analysis is the existence of regular non-chaotic orbits in the inner region. In
Section \ref{sec:4} we adiabatically change the parameters of the geodesic
orbits (imitating the dissipative behavior of gravitational waves) so that the
phase orbit crosses a chain of Birkhoff islands. Here we present a quantitative
result that relates the duration of the plateau with the various characteristics
of the central and of the orbiting bodies. We study the possible ways that this
crossing could take place and we get a crude estimate of the range of orbital
parameters of EMRIs that will eventually lead the system to pass through a
Birkhoff chain of islands of a strong resonance. In Section \ref{sec:5} we
analyze the evolution of the ratio of frequencies before, during, and after the
crossing of a resonance by the non-geodesic orbit and we obtain a plateau in the
ratio of the monitored frequencies during the crossing. We close our paper with
Section \ref{sec:6} by summarizing the physical conclusions drawn from our 
analysis. Finally in Appendix \ref{AppA} we give the relations between the
Boyer-Lindquist coordinates and the Keplerian elements, and in Appendix
\ref{AppB} we introduce a new set of variables, which are more appropriate for
our Fourier analysis.

\section{The Manko-Novikov metric}
\label{sec:2}

\subsection{Description of the MN spacetime}
\label{SubSec:ManNovIntro}

In order to demonstrate how the aforementioned new features appear and what are
their consequences, we are going to use the same type of non-Kerr metric which
was used by Gair et al \cite{Gair08}. It is a subclass of the multiparametric
family of spacetimes that Manko and Novikov constructed in 1992 \cite{Manko92}
as a generalization of the Kerr (and the Kerr-Newman) metric in an attempt to
describe the gravitational field of an arbitrary rotating and axially symmetric
isolated object. The corresponding construction, which is an exact solution of
the vacuum Einstein equations, was achieved by a non-linear superposition of the
Kerr spacetime with an arbitrary static vacuum Weyl field in a concise
analytical form (see \cite{90CastMank}). While the general solution of Manko and
Novikov does depend on an infinite number of parameters that are related to the
multipole moments of the gravitational field, the particular subclass that we
are going to study (the one studied in \cite{Gair08} as well) depends on three
parameters. Two of them determine the mass $M$ and the spin $S$ of the source of
the field, while the third one, $q$, is a dimensionless index that measures the
deviation of its quadrupole mass moment from the quadrupole moment of the
corresponding Kerr black hole (a Kerr metric that has the mass and the spin of
the new metric). Thus the first four non-zero mass and current-mass moments
$M_{l}, S_l$ (with $l\leq 3$) are fully characterized by these three parameters:
\be \label{MNMom}
\begin{array}{ll}
M_0 = M, &
S_1 = S, \\
M_2 = -M \left[\left( \frac{S}{M} \right)^2 +q M^2 \right], &
S_3 = -M \left[\left( \frac{S}{M} \right)^3 +2 q M^2 \left( \frac{S}{M} \right)
 \right],
\end{array}
\ee
while the higher order moments are in general higher order polynomials with
respect to $q$. We have chosen to present the moments in this way so that the
first part of each moment is merely the moment of the corresponding Kerr metric.
We remind that all multiple moments of a Kerr metric are characterized by two
parameters $M$ and $S/M$ and are given by the following concise formula:
\be
M_{l}+\imath S_{l}=M \left(\imath \frac{S}{M} \right)^l.
\ee

Having these in mind, we proceed to write the analytical expressions for the
metric functions of the particular subclass of Manko-Novikov metric (hereafter
called MN metric). As for all stationary, axially symmetric, and mirror
symmetric vacuum spacetimes the Weyl-Papapetrou line element for this metric is
\begin{eqnarray}
 ds^2 &=& -f(dt-\omega d\phi)^2+f^{-1} \left[
 e^{2\gamma} (d\rho^2+dz^2)+\rho^2 d\phi^2 \right]
\label{MNSpro}
\end{eqnarray}
where all metric functions $f,~\omega,~\gamma$ should be considered as functions
of the prolate spheroidal coordinates $x,y$ (the coordinates $\rho,z$ are the
corresponding cylindrical coordinates which could be expressed as functions of
$x,y$ as well). Thus
\be
\rho=k \sqrt{(x^2-1)(1-y^2)},\quad z=k x y
\ee
and
\begin{subequations}
\begin{eqnarray}
 f &=& e^{2 \psi}\frac{A}{B}, \label{ffunc} \\
 \omega &=& 2 k e^{-2 \psi}\frac{C}{A}-4 k \frac{\alpha}{1-\alpha^2}, \\
 e^{2 \gamma} &=& e^{2 \gamma^\prime}\frac{A}{(x^2-1)(1-\alpha^2)^2},
 \label{fexpgam} \\
 A &=& (x^2-1)(1+a~b)^2-(1-y^2)(b-a)^2,\label{fA} \\
 B &=& [(x+1)+(x-1)a~b]^2+[(1+y)a+(1-y)b]^2,\label{fB} \\
 C &=& (x^2-1)(1+a~b)[(b-a)-y(a+b)] \nonumber \\
   &&+ (1-y^2)(b-a)[(1+a~b)+x(1-a~b)], \\
 \psi &=& \beta \frac{P_2}{R^3}, \label{fC}\\
 \gamma^\prime &=& \ln{\sqrt{\frac{x^2-1}{x^2-y^2}}}+\frac{3\beta^2}{2 R^6}
 (P_3^2-P_2^2) \nonumber \\ &+& \beta \left(-2+\displaystyle{\sum_{\ell=0}^2}
 \frac{x-y+(-1)^{2-\ell}(x+y)}{R^{\ell+1}}P_\ell\right), \label{fgampr}\\
 a &=& -\alpha \exp {\left[-2\beta\left(-1+\displaystyle{\sum_{\ell=0}^2}
 \frac{(x-y)P_\ell}{R^{\ell+1}}\right)\right]}, \label{fa}\\
 b &=& \alpha \exp {\left[2\beta\left(1+\displaystyle{\sum_{\ell=0}^2}
 \frac{(-1)^{3-\ell}(x+y)P_\ell}{R^{\ell+1}}\right)\right]}, \label{fb}\\
 R      &=& \sqrt{x^2+y^2-1}, \label{fR}\\
 P_\ell &=& P_\ell (\frac{x~y}{R}), \label{fLegA}
\end{eqnarray}
\end{subequations}
where $P_\ell(z)$ denotes the Legendre polynomial of order $l$ given by
\be
P_\ell(z)=\frac{1}{2^\ell \ell!}
\left(\frac{d}{dz}\right)^\ell(z^2-1)^\ell.
\label{fLeg}
\ee 
The three parameters $k,\alpha,\beta$ that appear in the formulae above are the
three parameters that characterize the metric and are related to the mass $M$,
the spin $S$, and the quadrupole deviation $q$ through the following
expressions:
\begin{equation}
\begin{array}{r}
\alpha=\frac{-M+\sqrt{M^2-(S/M)^2}}{(S/M)},
\end{array}
\begin{array}{c}
k=M\frac{1-\alpha^2}{1+\alpha^2},
\end{array}
\begin{array}{l}
\beta=q \left( \frac{1+\alpha^2}{1-\alpha^2} \right)^3.
\end{array}
\label{freepar}
\end{equation}
The formula for $\alpha$ has been written as a function of $M$ and $(S/M)$ (both
of them have dimensions of mass), which are the physical parameters of the
central object that are common in the MN metric and the corresponding Kerr
metric. In \cite{Gair08} the formula for $\alpha$ is written as a function of
the dimensionless spin parameter $\chi=S/M^2$. We use the parameter $\chi$ later
in our numerical examples for simplicity.

The Kerr metric is a limiting case of the MN metric for $q=0$. A non-zero $q$
parameter distinguishes the two types of metrics. At this point we should note
that the MN metric is the general solution described in \cite{Manko92} with all
$a_i$ parameters set to zero except of $a_2$ which is the parameter $\beta$ that
shows up in the metric functions. Also, the $-2$ term in Eq.~(\ref{fgampr}) has
been moved outside the sum (c.f.,~Eq.~(3h) of \cite{Gair08}) so as to avoid any
confusion.

According to the thorough analysis of \cite{Gair08} the MN metric is not a black
hole solution, in agreement with the no-hair theorem. The central singularity is
surrounded by a  horizon which is broken along the equator by a circular line
singularity. As is the case with the Kerr metric, the MN spacetime possesses an
ergoregion (the region between the static limit $g_{tt}=0$ and the event horizon
$(g_{xx})^{-1}=0$) in the form of lobes that surround the horizon. Moreover a
region of closed time-like curves (where $g_{\phi \phi} <0$) is present (when
$q \neq 0$) in the form of lobes that overlap with the ergoregion. In all cases
we have investigated the region of permitted motion is not overlapping with the
region of the closed time-like curves, even though their borders may touch each
other.

Closing this general description of the MN metric we should emphasize once more
that this metric is an exact vacuum solution that can be continuously turned
into a Kerr metric by setting $q=0$. Thus the MN metric is a good candidate to
describe the field of a stationary axisymmetric and isolated spinning object
(the multipole moments of which differentiate the corresponding metric from a
Kerr one) outside some central region where singular behavior of the metric
functions shows up.

\subsection{Geodesics in the MN spacetime}
\label{SubSec:MNgeodesics}

Working in Lagrangian formulation, the geodesic orbits in the MN metric are
described as equations of motion of the following Lagrangian
\begin{equation}
L=\frac{1}{2}~\mu ~g_{\mu\nu}~ \dot{x}^{\mu} \dot{x}^{\nu}
\label{LagDef}
\end{equation}
where $\mu$ is the rest mass of the orbiting body and
$\dot{~}\equiv\frac{d}{d\tau}$ ($\tau$ denotes the proper time along the orbit:
$(d\tau)^2=-(ds)^2$). Such a Lagrangian of purely kinetic form is invariant
along the orbit. Its invariance is related to the fact that the rest mass of the
orbiting test body is a conserved quantity ($p^\nu p_\nu=-\mu^2$), and thus the
value of $L$ is $-\mu/2$. Due to stationarity and axisymmetry of the MN metric
there are two more integrals of motion, respectively; the specific energy
\begin{equation}
 E=-\frac{p_t}{\mu}=f(\dot{t}-\omega\dot{\phi})
 \label{EnIn}
\end{equation}
and the specific $z$-component of angular momentum of the orbiting test body
\begin{equation}
 L_z=\frac{p_\phi}{\mu}=f\omega(\dot{t}-\omega\dot{\phi})+\rho^2 \dot{\phi}/f
 \label{AnMoIn}
\end{equation}
(``specific'' means per unit rest mass). By suitable linear combinations of
these integrals of motion one yields the following expressions for $\dot{\phi}$
and $\dot{t}$:
\bea
\dot{\phi}&=&\frac{f}{\rho^2} (L_z-\omega E),
\label{phidot} \\
\dot{t}&=&\frac{\omega f}{\rho^2}\left[ L_z + \omega E
\left(\frac{\rho^2}{\omega^2 f^2} -1 \right) \right].
\label{tdot}
\eea
The two remaining equations of motion that determine $\rho(\tau)$ and $z(\tau)$
are thus sufficient to fully describe a geodesic orbit in such a spacetime; the
$\phi$ and $t$ coordinates are subsequently obtained by direct integration of
Eqs.~(\ref{phidot},\ref{tdot}).

Substituting the above expressions in the Lagrangian of Eq.~(\ref{LagDef}) and
taking into account the constant value of the Lagrangian itself, we obtain the
following constraint between the remaining four coordinates ($\rho,z,\dot{\rho},
\dot{z}$) of phase space
\be
\frac{1}{2}(\dot{\rho}^2+\dot{z}^2) + V_{\textrm{eff}}(\rho,z)=0,
\ee
where
\be
V_{\textrm{eff}}(\rho,z)=\frac{1}{2} e^{-2\gamma}
\left[
f-E^2+\left(\frac{f}{\rho}(L_z-\omega E) \right)^2
\right]
\label{Veff}
\ee
plays the role of an effective 2-dimensional potential. Note that this
expression is somewhat different from the expression for the effective potential
in Eq.~(13) of \cite{Gair08}; the present expression is written so as to
resemble better the newtonian analogue of energy conservation in potential
wells. The curve along which $V_{\textrm{eff}}=0$ determines the region of
allowed orbits in the polar plane (the $(\rho,z)$ plane that rotates along with
the orbiting body). The orbits can only move in the interior of such a curve,
since the effective potential is negative inside it (if we take into account
the third spatial coordinate $\phi$ as well it is actually a toroidal space-like
surface centered at the central singularity). Whenever an orbit reaches the
$V_{\textrm{eff}}=0$ curve both velocities $\dot{\rho},\dot{z}$ become equal to
zero. So we will denote this curve as the curve of zero-velocity (CZV).

\begin{figure}[htp]
\centerline{\includegraphics[width=36pc] {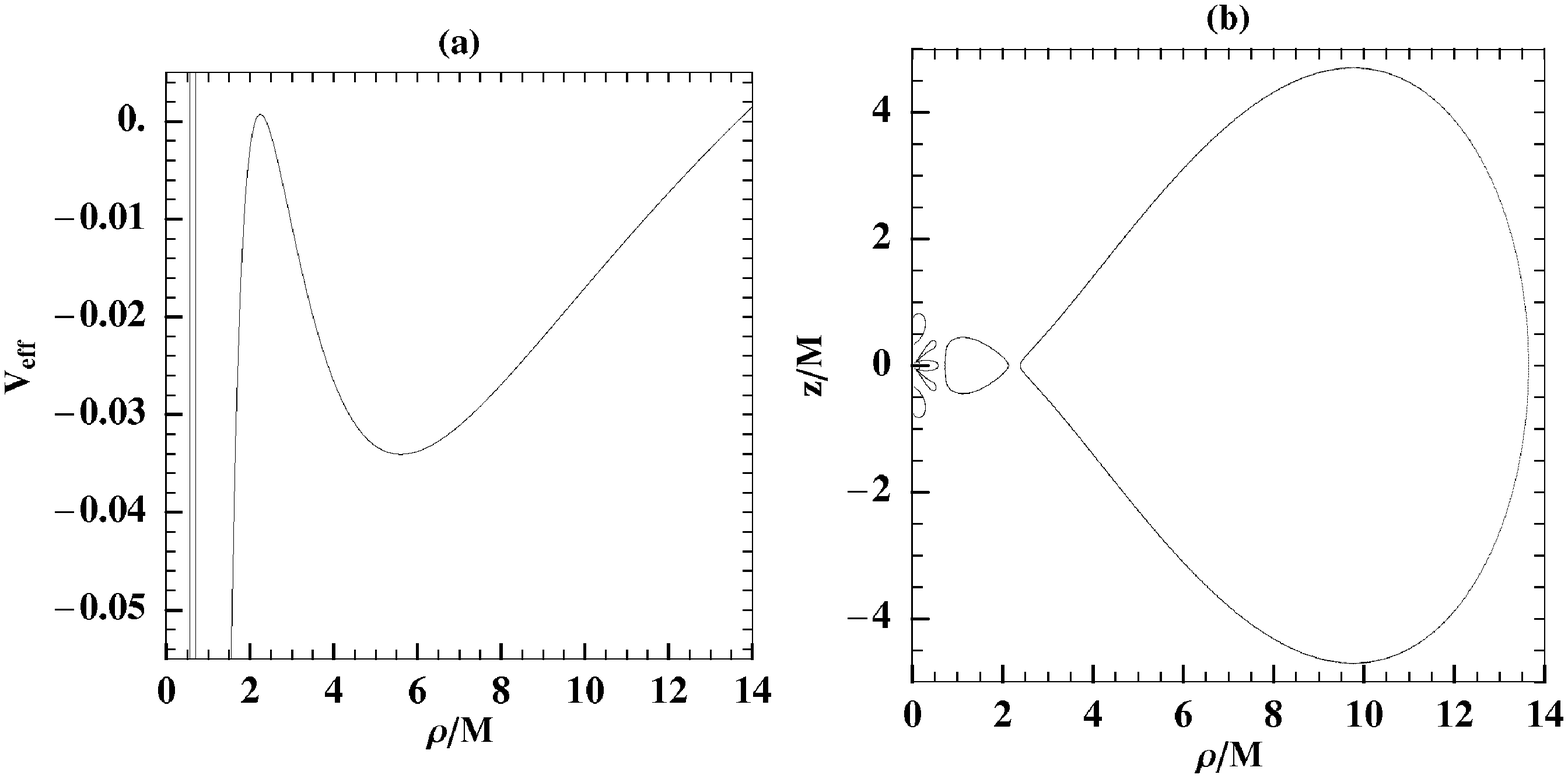}}
 \caption{(a) The shape of the 2-dimensional well $V_{\textrm{eff}}(\rho,z)$
 near  $V_{\textrm{eff}}(\rho,z)=0$ along the line $z=0$ for the MN metric with
 parameters $\chi=0.9,~q=0.95$ and orbital parameters $E=0.95, L_z=3 M$. (b) The
 CZVs ($V_{\textrm{eff}}(\rho,z)=0$) on the $(\rho,z)$ plane for the same set of
 parameters with (a).
}
 \label{fig:potential}
\end{figure}

In general the effective potential of the MN metric forms various distinct wells
in the $(\rho,z)$ polar plane (see Fig.~\ref{fig:potential}b). For $E<1$ there
are only bound orbits, since then the potential wells form CZVs that are closed
and are not extending to infinity. As explained in \cite{Gair08} for a prolate
perturbation of Kerr metric ($q < 0$) for a range of $E,L_z$ values there is a
distinct outer well and a lobe-like well which is connected to the horizon
within which we get plunging orbits. These two wells can be connected for a
suitable set of orbital parameters. On the other hand for oblate perturbations
of the Kerr metric ($q>0$) a new well shows up near the central region, in which
a new set of bound non-plunging orbits live. Therefore, in the latter case there
is a pair of closed CZVs, an inner one and an outer one
(Fig.~\ref{fig:potential}), which again can merge for a suitable range of
orbital parameters. It should be noted that the outer well is a shallow one that
resembles a lot the potential well of the Kerr metric. However the new inner
well is a much deeper one with intricate shape (part of this well is shallow but
there is also a very deep basin in it). Except from the regions of the
non-plunging orbits, in Fig~\ref{fig:potential}b the reader can discern 5 more
lobe-like regions of permitted motion which extend from the event horizon
($\rho=0$) to $\rho\lesssim 0.7 M$. These regions contain plunging orbits, a
fact that excludes them from the purpose of our paper, which is to study the
implications of non-integrability in bound non-plunging orbits.

The lower the energy, the higher is the effective potential
and the two distinct CZVs (if they are distinct) move further apart. The effect
of $L_z$ is the opposite of $E$: the higher the $L_z$, the CZVs move further
apart. 
Finally the effect of the source's spin is to bring the two CZVs further apart
for lower values of $S/M$. The last conclusion comes mainly from the fact that
lower $S/M$ values render the function $\omega$ in the metric less important;
hence the effect of the $L_z$ term in Eq.~(\ref{Veff}) effectively grows. The
inner well moves closer to the central region and eventually disappears, when
the ``anomalous quadrupole'' $q$ goes to zero from positive values. On the other
hand the outer well does not move or change its shape substantially for a wide
range of $q$ values.

Furthermore, the distinct inner well (the one that arises for positive $q$
values) is very close to the central region where anomalous regions are located.
E.g.~the ergoregion can extend into the interior region of the inner well. This
case will be examined carefully later on when we analyze the orbits in this
region. Fortunately the region of closed time-like curves remains outside the
inner well or it merely touches its boundary, at least for the range of
parameters which we have analyzed. We want to ensure that the orbits studied
do not enter the region of closed-timelike curves; if that happened it would
lead us to odd behavior that would have no clear physical meaning.

\section{The geodesic orbits in MN compared to Kerr}
\label{sec:3}

\subsection{The theoretical basis}
\label{sec:3.1}

In order to present the new observable features that arise in the MN spacetime
(as well as in any other similar non-Kerr spacetime) in contrast with the Kerr
spacetime, we will describe in brief some basic theorems of dynamical systems
that apply in the case of geodesic orbits in the MN spacetime.

Dynamical systems that are described by a Lagrangian function like
(\ref{LagDef}), can be described through a Hamiltonian function as well by
simply applying a Legendre transformation on the corresponding Lagrangian. The
Hamiltonian function that describes geodesic orbits in a MN metric (as well as
in the corresponding Kerr one, if one sets $q=0$) is
\be
H=p_\mu \dot{x}^\mu-L=\frac{1}{2 \mu} g^{\mu \nu} p_\mu p_\nu,
\ee
where instead of the velocities $\dot{x}^\mu$ (with respect to proper time)
we use the corresponding momenta $\displaystyle p_\mu=
\frac{\partial L}{\partial\dot{x}^\mu}=\mu g_{\mu\nu}\dot{x}^\nu$
to obtain the final Hamiltonian form.

The specific Hamiltonian that describes geodesic orbits in a MN spacetime, which
includes the Kerr spacetime as a special case, has no dependence on the
coordinates $t$ and $\phi$, due to stationarity and axisymmetry of the spacetime,
respectively. Thus the corresponding momenta $p_t$ and $p_\phi$ are conserved,
that is $p_t$ and $p_\phi$  are integrals of motion. These conserved momenta are
nothing but the quantities $-\mu E$ and $\mu L_z$, respectively, that we used
earlier. Furthermore, since $dH/d\tau=0$ (autonomous Hamiltonian), $H$ itself is
a third integral of motion, namely $H=-\mu$. The first two integrals of motion
can be used to reduce the number of degrees of freedom from 4 to 2; for example
one could use only the coordinates $\rho,z$ to fully describe this 2-dimensional
system. The value of the rest two coordinates $(\phi(\tau),t(\tau))$ along the
geodesic orbit could be inferred afterwards by simply integrating
Eqs.~(\ref{phidot},\ref{tdot}), as discussed in the previous section. Thus the
phase orbits lie on a 3-dimensional hypersurface of the corresponding
4-dimensional phase space, because of the conservation of $H$.

Now, if such a Hamiltonian system of 4 degrees of freedom is an integrable one,
that is there is one more integral of motion $I_4$ which is independent from the
previous ones and is in involution with them ($\{I_4,I_i \}=0$, where $I_i$
represents the first 3 integrals of motion and $\{\cdot,\cdot \}$ represents a
Poisson bracket), then the phase orbits lie on 2-dimensional surfaces. The Kerr
spacetime is an example of such an integrable case with the Carter constant
being this new extra integral of motion. This is exactly what makes the Kerr
spacetime a very distinct member of the broad family of stationary, axisymmetric
and asymptotically flat spacetimes. Moreover the natural formation of Kerr black
holes through gravitational collapse of astrophysical compact bodies according
to the no-hair theorem, renders the Kerr spacetime a unique highly symmetric
case of extreme astrophysical interest that has been extensively studied for a
long time \cite{bookKerr,Chandra}.

On the other hand any kind of generic perturbation of a Kerr metric that
maintains the stationarity and axisymmetry, most probably does not possess a
4-th integral of motion, as in Kerr. This new spacetime could arise from a
non-conventional astrophysical process that leads to other types of
ultra-compact objects, or it could simply be produced by an axisymmetric
distribution of matter around a Kerr black hole (i.e. an accretion disc).

In an integrable autonomous system of 2 degrees of freedom, like the one
describing the geodesic orbits in a Kerr metric, the bound orbits lie on
2-dimensional tori in the 4-dimensional phase space. The values of the integrals
of motion (the Hamiltonian and the Carter constant) fully characterize these
tori; the tori that correspond to different Carter-constant values are nested
within each other, while the Hamiltonian value defines the overall scale of the
phase space. If one considers a 2-dimensional surface that cuts through a
foliage of such tori (called Poincar\'{e} surface of section), each torus
defines a closed curve on this surface (see e.g.~\cite{Lichtenberg92}). This
curve is called an invariant curve. Each torus corresponds to a characteristic
pair of frequencies (one for each angle variable). Not only the two frequencies,
but also the ratio between them varies continuously from one torus to the next.
If the ratio of the frequencies is an irrational number, a phase-orbit
continuously winds around its corresponding torus covering densely the surface
of the torus. This kind of orbit is called quasiperiodic. A quasiperiodic orbit
goes repeatedly through a surface of section defining a succession of points
which eventually cover densely the corresponding invariant curve on the surface
of section (called quasiperiodic invariant curve). In the special case where the
ratio of frequencies is a rational number $n/m$ ($n,~m \in \mathbb{N}$) the
phase-orbit repeats itself after $m$ windings; then the orbit is periodic and
the corresponding torus is called resonant. In this case the resonant invariant
curve consists of an infinite number of $m$-multiplets of periodic points. Each
$m$-multiplet represents an $m$-multiple periodic orbit.

\begin{figure}[htp]
 \centerline{\includegraphics[width=14pc] {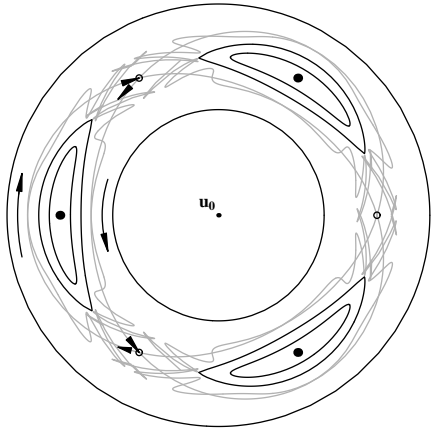}}
 \caption{A schematic representation of a surface of section of a nonintegrable
 system of 2 degrees of freedom. On the surface there are depicted two KAM
 curves; one outside and one inside the chain of islands. The particular chain
 of islands in between consists of three islands ($m=3,~n=1$), each one arising
 around each of the three stable points (shown as filled circles). Each island
 consists of a whole set of KAM curves nested within each other. Between the
 successive islands there are three unstable points (shown as open circles) from
 which the asymptotic curves (gray curves) emanate surrounding the islands by a
 thin chaotic layer (here magnified). The arrows indicate the flow around the
 leftmost island. The $\mathbf{u}_0$ indicates the central fixed point around
 which the KAM curves and the Birkhoff islands are formed.
 }
 \label{FigSurSecNI}
\end{figure}

According to the Kolmogorov-Arnold-Moser (KAM) theorem \cite{KAM}, if any
integrable system gets perturbed (without altering its dimensionality) most of
its tori are deformed but they are not destroyed. These tori are called KAM tori.
Thus the corresponding surface of section looks very much like the surface of
section of the corresponding integrable system. The quasiperiodic invariant
curves in nonintegrable systems are called KAM curves. This is exactly the
behavior exhibited by most of the geodesics in the MN spacetime, at least for
bound orbits in the outer allowed region (see \cite{Gair08}). However, there are
qualitatively new characteristics that are related with the resonant tori of a
perturbed integrable system. According to the Poincar\'{e}-Birkhoff theorem
\cite{PB}, when a system deviates slightly from an integrable one, the resonant
invariant curve disintegrate and only a finite even number ($2~k~m$, where
$k \in \mathbb{N}^* $) of the periodic points of period $m$ survive. This means
that from the resonant torus only $2 k$ periodic orbits survive. Half of the
surviving periodic points are stable while the rest are unstable. If we imagine
a closed curve which passes from all the surviving points of the disintegrated
resonant curve, then the stable and the unstable periodic points interchange
along that closed curve forming the Birkhoff chain. Around each stable periodic
point there is a set of nested KAM curves forming an island of stability (see
Fig.~\ref{FigSurSecNI}). A phase-orbit of such a resonant case visits all the
$m$ islands of the $n/m$-resonance, moving successively to the next $n$-th
island along the imaginary aforementioned closed curve at every winding, forming
eventually the KAM curves inside every island. The interesting feature of these
islands, that we exploit in our paper, is that every regular (non-chaotic) orbit
that belongs to a chain of islands is characterized by the same rational ratio
of frequencies (characteristic of the particular chain of islands), no matter on
which KAM curve inside the chain of islands it belongs. This property is not
shared by the non-resonant KAM curves, since the ratio of frequencies on them is
irrational and it varies smoothly from one KAM curve to another.

Finally, there is a region around the Birkhoff chain of islands that consists of
chaotic orbits. These chaotic orbits appear mainly in the neighborhood of the
unstable points of the Birkhoff chain (see Fig.~\ref{FigSurSecNI}). The chaotic
orbits arise from the asymptotic curves of the unstable periodic orbits, which
are forced to follow a very complicated pattern of multiple intersections when
they approach each other. The intersections of the asymptotic curves are called
homoclinic points and the corresponding orbits are called homoclinic orbits. The
chaos produced by the homoclinic orbits is called homoclinic chaos. As long as
the deviation from the integrable case is small the chaotic areas on a surface
of section are restrained on a very thin ring, surrounding the Birkhoff islands
of stability, which is thicker near the unstable points. For higher deviations
from an integrable case the KAM curves intervening between two different
resonances are destroyed. Then two Birkhoff chains which were initially isolated
by intermediate KAM curves can communicate (resonance overlap)
\cite{ResOver,Chirikov79,Contop02}, i.e.~the asymptotic curves that emanate from
unstable points of different chains intersect each other at the so-called
heteroclinic points, forming heteroclinic orbits. These orbits produce the
so-called heteroclinic chaos. The chaotic behavior in a resonance overlap is
stronger than in the homoclinic one.

It should be emphasized that the $q$-parameter that characterizes the deviation
of the MN metric from the corresponding Kerr metric is not an absolute measure
of the deviation of the corresponding geodesics, since different regions of the
MN spacetime have more or less deformed metric components (with respect to their
Kerr values). For example the greater the distance from the central point of the
field, the smaller is the effect of the perturbed moments of the spacetime,
since there the spacetime is mainly characterized by its lowest few moments;
there the higher multiple moments are almost unimportant for the shape of the
geodesic orbits. Therefore one anticipates that the system of MN geodesic orbits
should be more organized in its outer region, in contrast with its more chaotic
inner region.

\subsection{Rotation number}
\label{sec:3.2}

When the perturbation parameter of a 2-dimensional nonintegrable system is
sufficiently small, the Birkhoff islands of stability are very thin and their
detection on a surface of section is quite cumbersome; fine tuning is needed to
find suitable initial conditions of orbits that develop into a chain of islands
on a surface of section. However, these islands can be detected by a more
sophisticated technique. The islands of stability lie around a resonant periodic
orbit which is characterized by a commensurate ratio of frequencies
$\nu=\omega_1/\omega_2=n/m$, where $\omega_1,~\omega_2$ are the fundamental
frequencies corresponding to the two angle variables while $n, m$ are integers.
This ratio characterizes not only the resonant periodic orbit of the island
(represented by a set of $m$ stable points on a surface of section), but all
the KAM orbits around them belonging to the particular chain of islands.
Although each distinct KAM orbit in an island is characterized by a different
pair of frequencies $\omega_1,~\omega_2$, all such KAM orbits are marked by the
same commensurate ratio $\nu$ with the central periodic orbit.

An index that has been used for detecting chaos in classical nonintegrable
systems of two degrees of freedom is the rotation number. This index computes
the ratio of the fundamental frequencies $\omega_1,~\omega_2$ and therefore
could be used to detect the islands of stability as well
\cite{Contop02,Laskar93,Voglis98}. In order to evaluate the rotation number we
first identify the fixed point $\mathbf{u}_0$, around which the KAM curves
(not the ones a Birkhoff island consists of) are formed (Fig.~\ref{FigSurSecNI})
creating a formation known as the main island of stability. In our case
$\mathbf{u}_0$ is the point that corresponds to the periodic orbit which crosses
the equatorial plane at only one point with $\dot{\rho}=0$ moving towards the
positive part of the $z$-axis (see Fig.~\ref{FigExtReg}a). It should be noted
though that this orbit is periodic with respect to its projection on the
$(\rho,z)$ plane, but if one considers the $\phi$ coordinate of the orbit as
well, the orbit is not necessarily periodic then. Now we define the position
vector of the $i$-th crossing point $\mathbf{u}_i$ of a phase orbit on a surface
of section to be
\be
\mathbf{r}_i=\mathbf{u}_i-\mathbf{u}_0,
\ee
that is its position with respect to $\mathbf{u}_0$. Next we compute the angles
$\theta_i\equiv angle (\mathbf{r}_{i+1},\mathbf{r}_i)$ between two successive
position vectors, the so-called rotation angles, and finally we calculate
their mean value for a large number (theoretically an infinite number) of
crossings, divided by $2\pi$. This number provides the so-called rotation number
$\nu_\theta$, i.e.
\begin{equation}\label{RotNu}
\nu_\theta=\lim_{N\rightarrow\infty}\frac{1}{2\pi N}\sum_{i=1}^N \theta_i
\end{equation}
and it measures the average fraction of a circle by which successive crossings
advance.

The rotation number usually appears to grow monotonically as long as we cross
KAM curves that surround the central fixed point $\mathbf{u}_0$. The strict
monotonicity is interrupted by KAM curves belonging to resonant islands of
Birkhoff chains surrounding the fixed point $\mathbf{u}_0$. Within a resonant
island the rotation number remains fixed at a constant rational value. This
rational number, as already stated, is characteristic of that island of
stability, since it is the outcome of the resonance between $\omega_1,~\omega_2$.
It should be also noted that in a chaotic region of a nonintegrable system the
rotation number fluctuates irregularly from point to point, so its behavior
appears to be smooth only in the region of regular orbits. For a slightly
perturbed integrable system the chaotic layers around the Birkhoff chains of
islands are so thin that the fluctuating behavior of the rotation number is
hardly observable.

The rotation number is not just a ``mathematical'' tool, it is actually an
``observable'' quantity. The frequencies $\omega_1$, $\omega_2$ can be derived
from the signal of the gravitational waves, through a Fourier analysis. By
monitoring the evolution of the frequencies encoded in a gravitational wave
signal coming from an EMRI, we monitor the rotation number and when we observe
stationarity of its value as time progresses we can infer that the corresponding
EMRI is evolving in a non-Kerr spacetime.

\subsection{The outer region}
\label{subsec:outer}

\begin{figure}[htp]
 \centerline{\includegraphics[width=31pc] {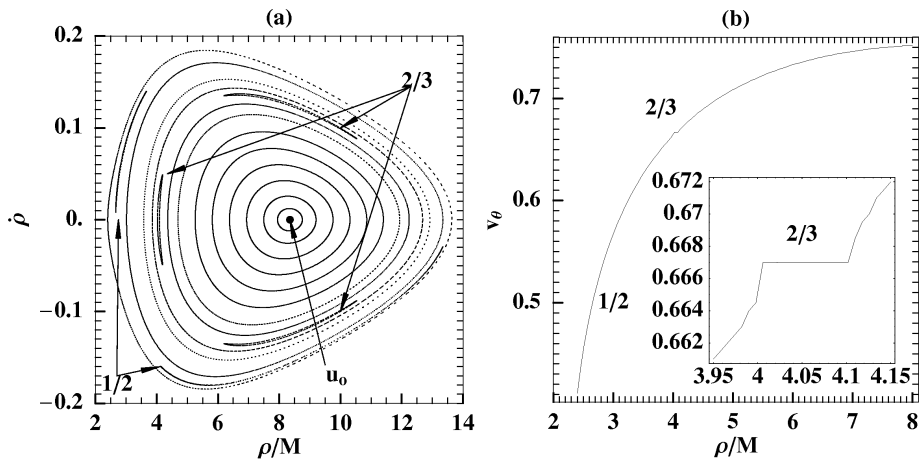}}
 \caption{ (a) The surface of section of the outer region on the
 $(\rho,~\dot{\rho})$ plane for the parameter set
 $E=0.95,~L_z=3 M,~\chi=0.9,~q=0.95$. $\mathbf{u}_0$ indicates the fixed point
 at the center of the main island. (b) The rotation number vs. $\rho$ along the
 line $\dot{\rho}=0$ of the surface of section presented in (a). Embedded in (b)
 is a detail of the rotation curve around the $2/3$-resonance.
 }
 \label{FigExtReg}
\end{figure}

As it was demonstrated in Fig.~\ref{fig:potential}, which was characterized by
a particular set of parameters of the MN spacetime and of the test particle,
there may be two separate regions where bound geodesic orbits are allowed to
develop, namely the inner region (closer to the central singularity) and the
outer region (farther from the central singularity). In this section we explore
the dynamical properties of the outer region by studying the surface of section
$z=0$ ($\dot{z}>0$) on the $\rho,~\dot{\rho}$ plane shown in
Fig.~\ref{FigExtReg}a. By inspection this surface of section seems to be filled
densely with KAM curves, and only two thin chains of islands of stability are
discerned; no visible sign of stochasticity is present. However, the existence
of these two Birkhoff chains, the one of multiplicity 2 (labeled as 1/2) and the
other of multiplicity 3 (labeled as 2/3), implies that the system is actually
nonintegrable and therefore chaos should be present \cite{Chirikov79,Contop02}.
In fact the surface of section should be densely filled with other Birkhoff
chains of islands as well, but their detection demands a very detailed scan of
the surface of section. Even the revealing of the two chains of islands which we
found would be a very hard task, if we hadn't employed the rotation number in
order to explore the fine details of the phase space that did not show up in the
coarse sweep of initial conditions performed by Gair et al \cite{Gair08}.

As we have already explained the rotation number is an appropriate tool to
detect the islands of stability, since all the regular orbits belonging to an
island share a common characteristic rational value of the rotation number. The
implementation of the rotation number in our case is presented in
Fig.~\ref{FigExtReg}b. The scan begins from the central point $\mathbf{u}_0$ and
goes inwards, towards $\rho=0$, along the line $\dot{\rho}=0$ on the surface of
section seen in Fig.~\ref{FigExtReg}a. The produced rotation curve is a
decreasing function of the distance from the center $\mathbf{u}_0$ of the main
island of the outer region. The curve appears to be strictly monotonic except
from a narrow plateau, a constant value of the rotation number within a
$\rho$-interval, lying near $\rho=4$ (see the magnified plot of the embedded
figure in Fig.~\ref{FigExtReg}b). This plateau is labeled by the value of the
corresponding rotation number $\nu_\theta=2/3$, and is related to the period-3
chain of islands of Fig.~\ref{FigExtReg}a.

\begin{figure}[htp]
 \centerline{\includegraphics[width=14pc] {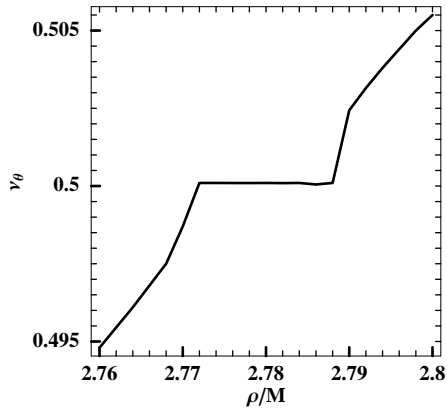}}
 \caption{
 The rotation curve along the line $\dot{\rho}=0.05$ which is crossing the
 $1/2$-resonance on the surface of section of Fig. \ref{FigExtReg}a.
  }
 \label{FigRes2}
\end{figure}

However no plateau appears in Fig.~\ref{FigExtReg}b for the period-2 island of
stability  ($1/2$-resonance), even though the rotation curve passes through the
value $\nu_\theta=1/2$. This can only mean that the island is very thin along
the line $\dot{\rho}=0$ and our step along the scanning line is not sufficiently
small to discern a corresponding plateau. Instead by moving along the line
$\dot{\rho}=0.05$, where the particular period-2 island is thicker, we find a
discernible plateau (Fig.~\ref{FigRes2}).

In the integrable Kerr case no such plateaus are expected to exist. According to
the analysis done in \cite{Schmidt02} for the bounded geodesic orbits in the
Kerr spacetime, the Hamilton function expressed in action angle variables does
not depend explicity on the angle variables and therefore the integrable Kerr
spacetime has no resonant islands. Thus the rotation curve in a Kerr case is a
strictly monotonic function of the distance from the center; the corresponding
resonances being presented simply by a set of periodic points that lie along a
single curve on a surface of section instead of a chain of islands with finite
thickness. This difference, along with the analysis of inspiraling orbits
presented in section \ref{sec:4}, is an effect that can be quantitatively
checked by the gravitational wave detectors.

Once again we note that other resonances corresponding to any rational value of
$\nu_\theta$, that are not depicted in Fig.~\ref{FigExtReg}, are so narrow that
they are really very difficult to be pinpointed on a surface of section.

\begin{figure}[htp]
 \centerline{\includegraphics[width=36pc] {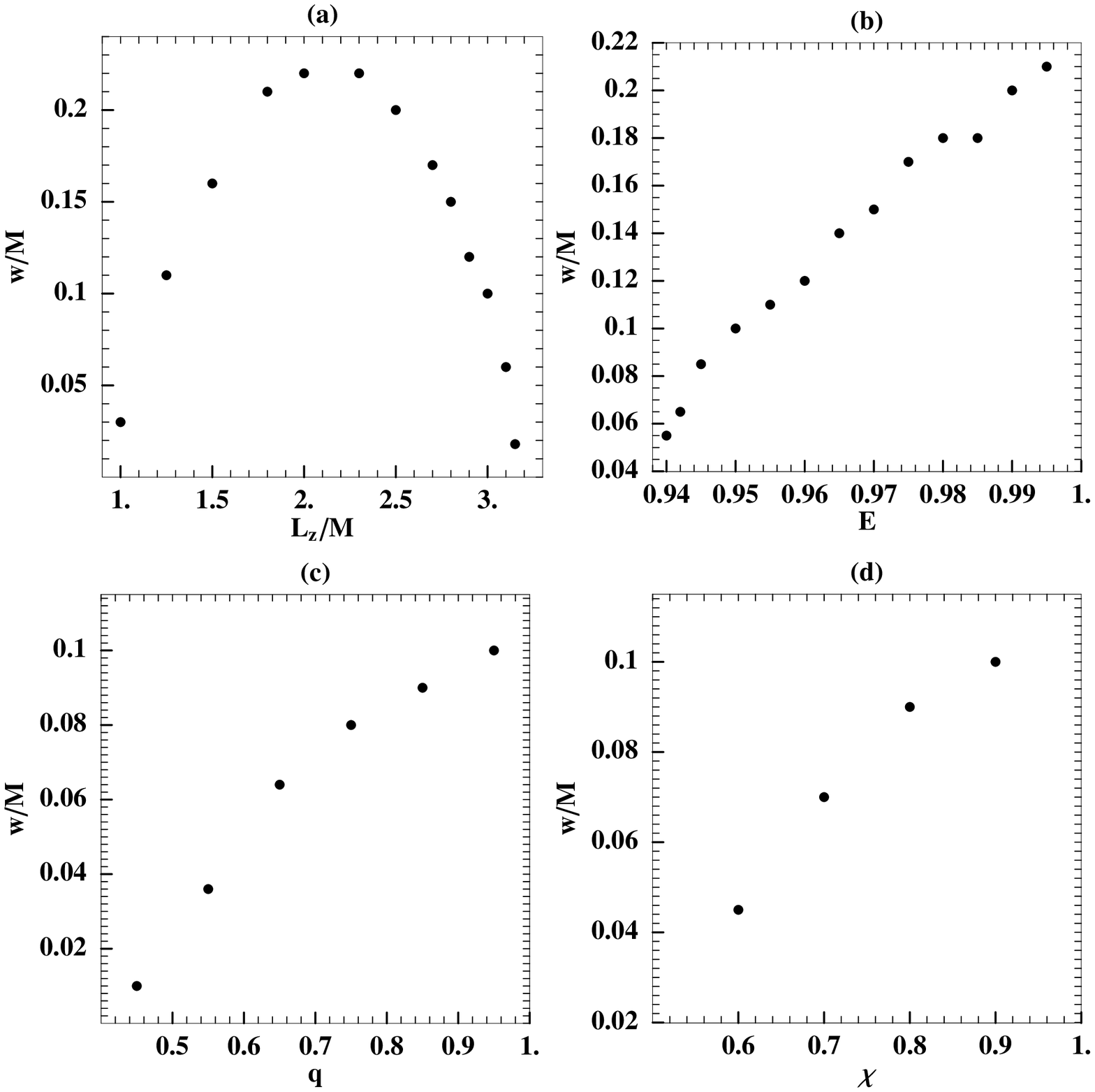}}
 \caption{The width $w$ of the period-3 leftmost island of stability along the
 line $\dot{\rho}=0$ (the thickest region) for different sets of the $E,~L_z,~q$,
 and $\chi$ parameters.
 (a) The $w$ vs.~the angular momentum $L_z$ when $E=0.95,~q=0.95,~\chi=0.9$.
 (b) The $w$ vs.~the energy $E$ when $L_z=3 M,~q=0.95,~\chi=0.9$.
 (c) The $w$ vs.~the quadrupole deviation $q$ when $L_z=3 M,~E=0.95,~\chi=0.9$.
 (d) The $w$ vs.~the spin parameter $\chi$ when $L_z=3 M,~E=0.95,~q=0.95$.
 Apparent non-smooth behavior of the plots is an artifact of the accuracy
 used to measure the width.
 }
 \label{FigWidth}
\end{figure}

\begin{figure}[htp]
 \centerline{\includegraphics[width=36pc] {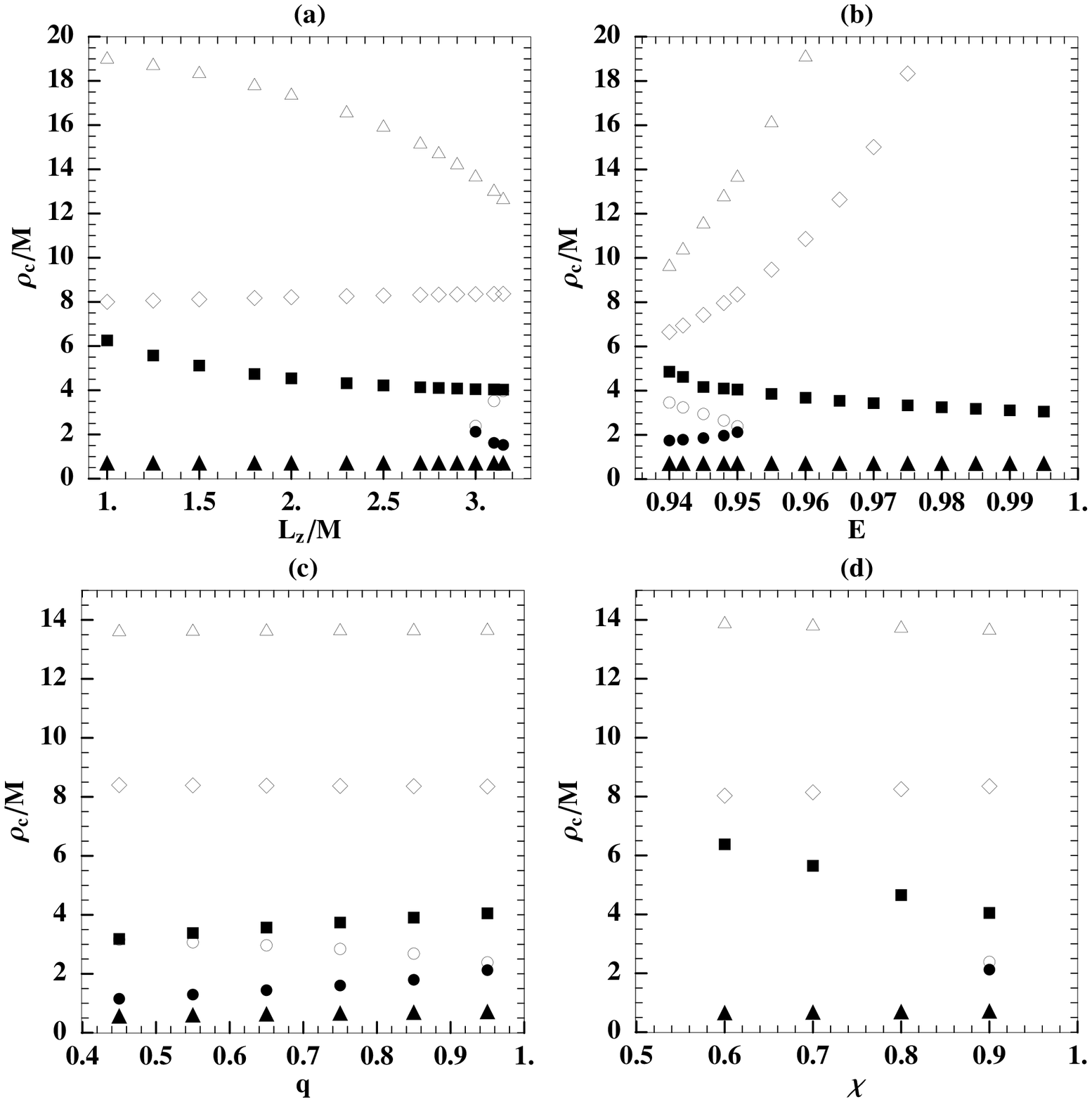}}
 \caption{ The radial positions $\rho_c$ of 6 different locations of interest
 along the line $\dot{\rho}=0$ on the surface of section $z=0$ for the same sets
 of parameters $E,~L_z,~q,~\chi$ as in Fig. \ref{FigWidth}. The diamonds
 represent the central point $\bf{u}_0$ of the main island, the black squares
 represent the center of the multiplicity-3 leftmost island, the open triangles
 and open circles represent the outer and inner boundary of the outer region
 respectively, while the black triangles and black circles represent the inner
 and outer boundary of the inner region respectively. When the open and the
 black circles merge the outer and the inner regions get connected through a
 neck.
  }
 \label{Fig:islandpositioncenter}
\end{figure}

As mentioned before the finite width of the resonant islands in a non-Kerr
metric, in contrast to the zero-width of the corresponding resonances of the
Kerr metric is the new feature that we propose to exploit. Thus the size and the
position of the islands are of great importance in our analysis. The size and
the position of the resonant islands depend on both the physical parameters of
the MN metric ($q$ and $\chi$) and the parameters of the orbit itself ($E$ and
$L_z$). For example if we keep the three parameters $E,\chi,q$ fixed and plot
the width $w$ of the period-3 island along the $\dot{\rho}=0$ line on the
surface of section $z=0$ as a function of the angular momentum $L_z$, we get
Fig.~\ref{FigWidth}a. On the other hand in Fig.~\ref{Fig:islandpositioncenter}a
we plot along the same line, for the same set of parameters, the positions
$\rho_c$ of the central point $\mathbf{u}_0$ of the main island (diamonds), the
center of the multiplicity-3 leftmost island (black squares), the outer and
inner boundary of the outer region (open triangles and open circles
respectively), the inner and outer boundary of the inner region (black triangles
and black circles respectively). The width $w$ of the island increases as we
increase $L_z$, till it reaches its maximum width when $L_z\approx 2 M$
(Fig.~\ref{FigWidth}a). Then it starts becoming thinner again. This behavior is
due to the fact that as we increase $L_z$ the potential well becomes more
compressed along the equator. Therefore as the particular island moves away from
the central point $\mathbf{u}_0$ it is compressed by the boundary of the well
(high $L_z$ values), while if the island moves close to the center, it is
finally compressed in the region around the central point $\mathbf{u}_0$ (low
$L_z$ values) (cf.~Figs.~~\ref{FigWidth}a,~\ref{Fig:islandpositioncenter}a).

If instead of varying the angular momentum we vary the energy $E$ and measure
the width $w$ and the positions $\rho_c$, we obtain Figs.~\ref{FigWidth}b and
\ref{Fig:islandpositioncenter}b respectively. Again as we increase the energy
the allowed region is expanded and thus the thickness of the island increases as
well. On the other hand there is a lower energy threshold below which the island
of the $2/3$-resonance is compressed to zero thickness due to shrinkage of the
whole CZV.

If we keep the energy and the angular momentum at their initial value
$E=0.95,~L_z=3 M$ and vary the quadrupole deviation $q$, the width $w$ of the
island and the positions $\rho_c$ change as shown in Figs.~\ref{FigWidth}c and
\ref{Fig:islandpositioncenter}c respectively. The $q$ parameter is actually the
parameter that controls the non-integrability of the system since for $q=0$ it
describes a Kerr metric which is integrable. Thus it is anticipated that as $q$
increases the width increases as well. Moreover the displacement of the
$2/3$-resonance with respect to the boundaries of the CZV works in favor of the
island's expansion as long as the island is away from the main center
$\mathbf{u}_0$. For example, as is shown in Fig.~\ref{FigWidth}c for $q$ below
$\simeq 0.4$ the thickness has been shrunk to almost zero, since the island has
moved close to the boundaries of the CZV (Fig.~\ref{Fig:islandpositioncenter}c).
It should be noted that this limiting value of $q$ has no universal character,
but corresponds to the specific values of the other parameters.

Finally the role of the spin parameter $\chi$ of the metric is qualitatively
similar to that of the angular momentum of the orbit. The island moves away from
the central point $\mathbf{u}_0$ as $\chi$ increases
(Fig.~\ref{Fig:islandpositioncenter}d), while the thickness of the island
increases as well (Fig.~\ref{FigWidth}d). At least in the case we have studied
(corresponding to the specific values for the other parameters) when
$\chi \simeq 1$ the island has not yet come very close to the boundaries of
allowed orbits (Fig.~\ref{Fig:islandpositioncenter}d); thus the island has not
been forced to shrink (Fig.~\ref{FigWidth}d).

For all the above range of parameters the period-3 island exists along with
other islands of stability from different Birkhoff chains. In general when the
Kerr spacetime is even slightly perturbed and the Carter constant is destroyed
(ceases to be a constant), the Birkhoff chains with their islands of stability
show up. The general behavior of the most prominent islands of resonance in a
slightly perturbed Kerr metric are expected to exhibit similar behavior with the
$2/3$-island for different ranges of the parameters $E,~L_z,~q,~\chi$ of the MN
metric, since the parameters $E$, $L_z$ of the orbit, and $\chi$ of the metric
play a similar role for any kind of orbit in a generically perturbed Kerr metric.
The $q$ parameter of the MN may be replaced by another set of one or more
parameters in a generic perturbed Kerr metric that will control the deviation of
the metric from the Kerr metric.

In order to numerically integrate the orbits and produce Figs.~\ref{FigExtReg}-
\ref{Fig:islandpositioncenter} we have applied a sixth order Runge Kutta
integration scheme with a constant step of integration $\delta s=10^{-1}$,
except for the cases in which an orbit reaches the surface of section; then the
integration step was gradually reduced to $\delta s=10^{-8}$ in order to get a
fine approximation of the surface of section at $z=0$. To ensure that the
numerical errors do not affect the evolution of the orbits and our conclusions
with regard to their characteristics, at every step we evaluate the Lagrangian
per unit test mass $L_n/\mu$ and compare it to its expected value $L/\mu=-1/2$.
The relative error of the integration $\Delta L/L=|(L_n-L)/L|$ after $10^3$
crossings through the $z=0$ surface of section did not exceeded the order of
$\Delta L/L=10^{-10}$. In order to get a smaller relative error we tried to use
an even smaller step of integration for these orbits. This was time consuming
though, without improving substantially any quantitative information we gained
with a coarser integration step; therefore we decided to keep the initial
$\delta s=10^{-1}$ step.

\subsection{The inner region}
\label{subsec:inner}

\begin{figure}[htp]
 \centerline{\includegraphics[width=32pc] {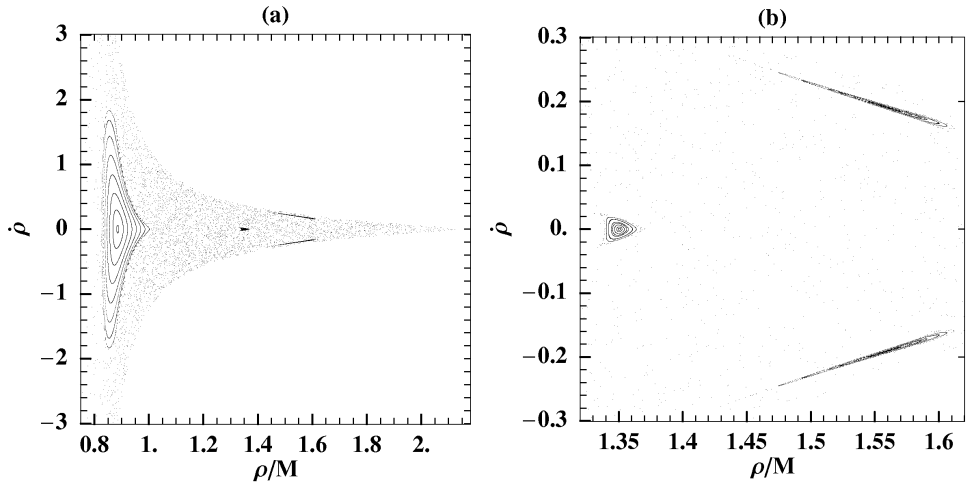}}
 \caption{ (a) The surface of section of the inner region on the
 $\rho,~\dot{\rho}$ plane for the parameter set
 $E=0.95,~L_z=3 M,~\chi=0.9,~q=0.95$. (b) A detail of (a) showing islands of
 stability surrounding a period-3 orbit.
 }
 \label{FigIntReg}
\end{figure}

In contrast to the outer region, which contains mainly regular orbits, the inner
region displayed in Fig.~\ref{FigIntReg} has a more complicated structure. On
the left side of Fig.~\ref{FigIntReg}a there is a period-1 island of stability.
Around the center of the same figure there are also three islands of stability
belonging to a period-3 chain (these islands are depicted in detail in
Fig.~\ref{FigIntReg}b). The rest of the inner region seems to be occupied by
chaotic orbits, in accordance with the findings of Gair et al \cite{Gair08}. A
domain of the surface of section that is mainly occupied by chaotic orbits
surrounding a main island of stability, like the region of
Fig.~\ref{FigIntReg}a which is filled with scattered dots, is often described
as a chaotic sea in nonlinear dynamics literature (see e.g.~\cite{Contop02}).

To produce Fig.~\ref{FigIntReg} we applied again a sixth order Runge Kutta
integration scheme but now we used an adaptive step of integration which varied
from $\delta s=10^{-4}$ to $\delta s=10^{-8}$ in a way that the relative error
did not exceeded $\Delta L/L=10^{-8}$. In order to avoid problems with the
static limit, which is extended inside the inner region, we analytically
eliminated the zeroes of function $A$  (Eq.~(\ref{fA})) from all the metric
components $g_{\mu\nu}$ for which $A$ appears in the denominator, like $g_{\phi
\phi}$.

Oddly enough Fig.~\ref{FigIntReg}a seems not to be in total agreement with
Fig.~7 of \cite{Gair08}. The region which in our Fig.~\ref{FigIntReg}a is
occupied by a main island of stability, in Fig.~7 of \cite{Gair08} seems to be
visited by a chaotic orbit coming from the chaotic sea. In fact, in a system of
2 degrees of freedom chaotic orbits surrounding an island of stability cannot
enter the island itself, because the KAM curves of the island in such systems
block such entrances.

\begin{figure}[htp]
 \centerline{\includegraphics[width=29pc] {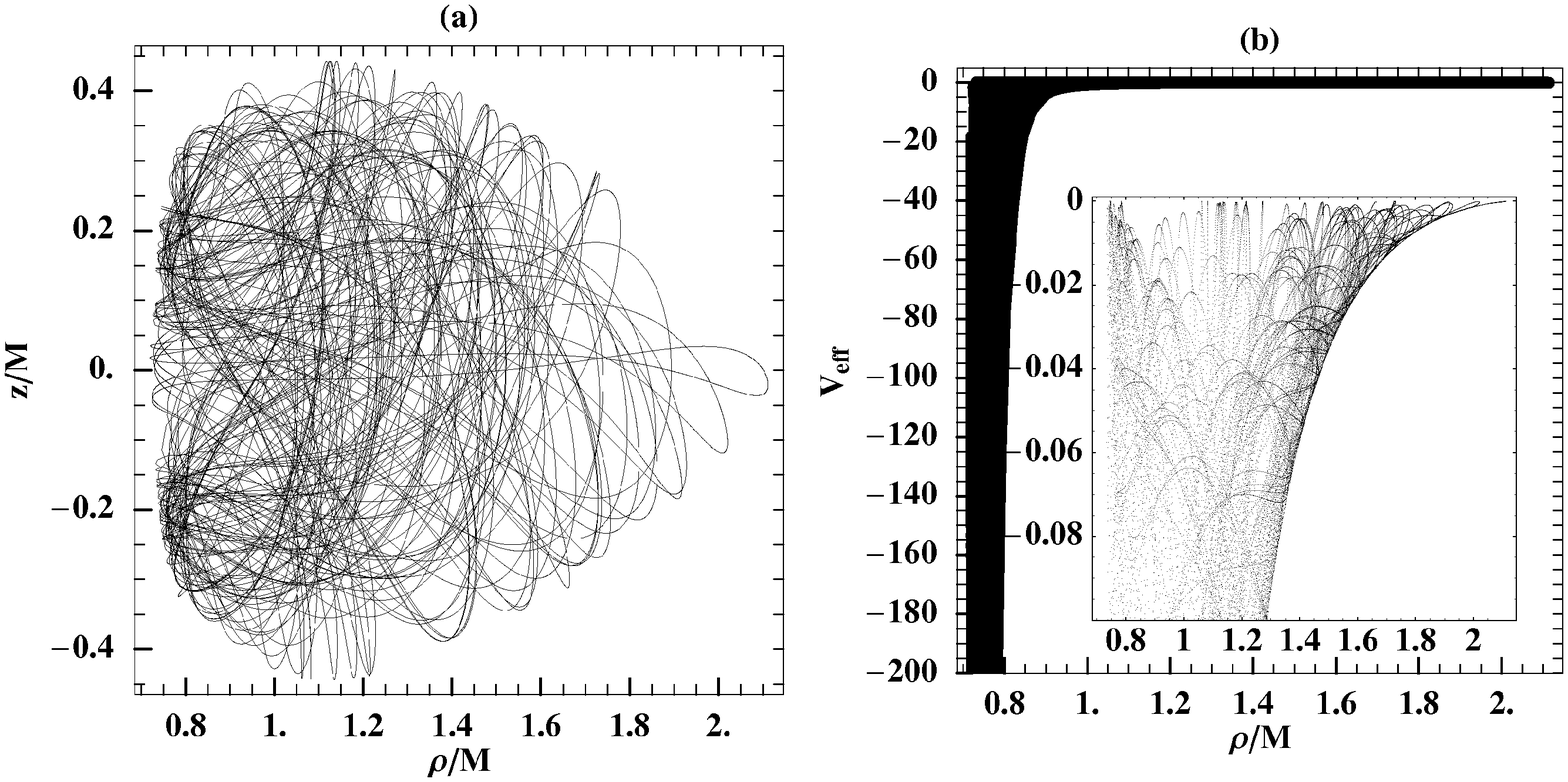}}
 \centerline{\includegraphics[width=29pc] {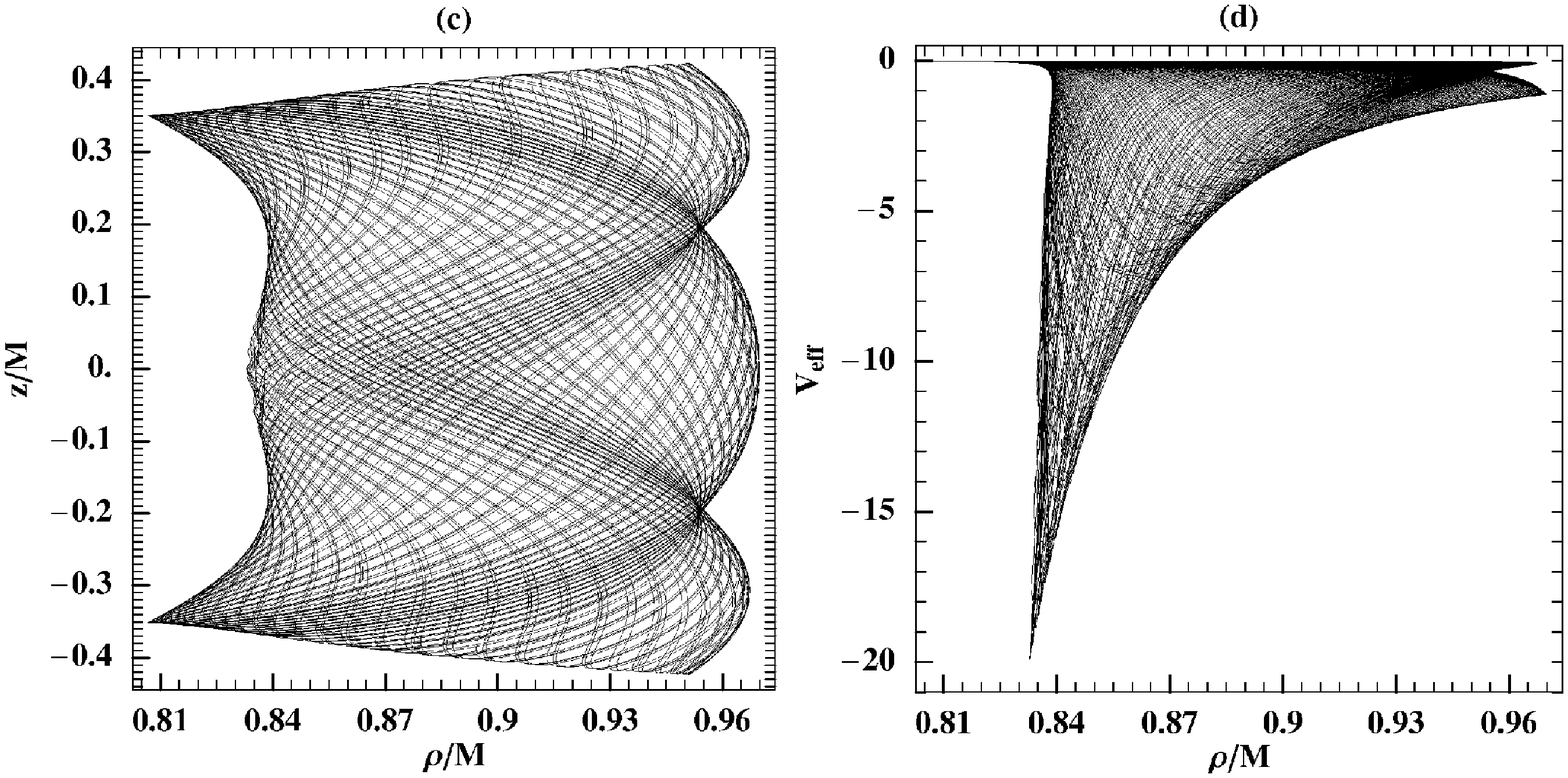}}
 \centerline{\includegraphics[width=29pc] {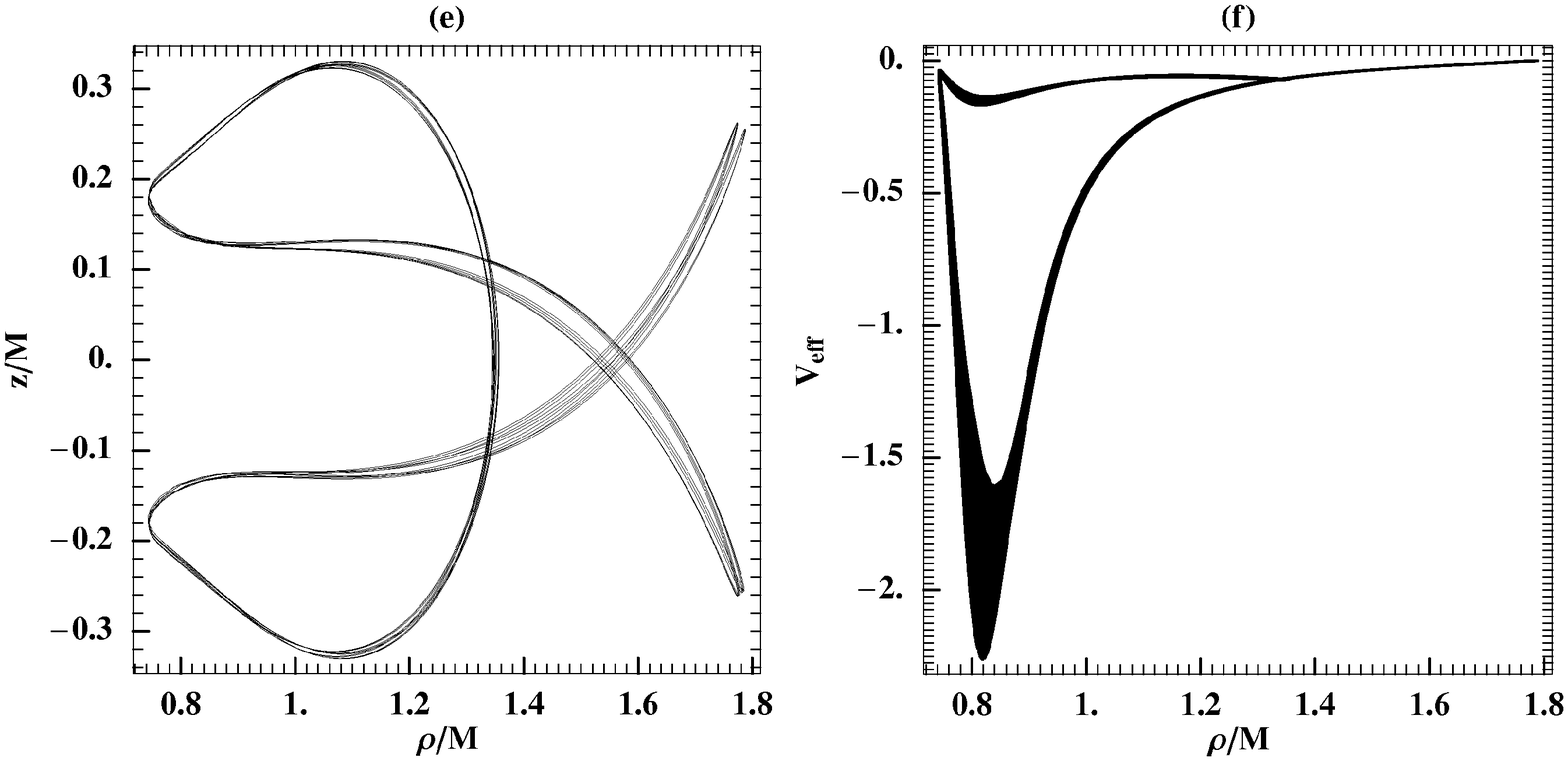}}
 \caption{ In the left column various bound orbits of the inner region
 (Fig.~\ref{FigIntReg}) are  shown as projections on the $(\rho,z)$ plane. The
 projected orbits are: (a) a chaotic orbit, (c) a regular orbit of the main
 island, (e) a regular orbit of the 2/3-resonance. The right column shows the
 corresponding evolution of the value of $V_{\textrm{eff}}$ along the orbit as a
 function of the instantaneous $\rho$ coordinate. The embedded figure in (b)
 shows a detail near the upper edge of $V_{\textrm{eff}}$
 ($V_{\textrm{eff}} \simeq 0$).
}
\label{FigVeffOrb}
\end{figure}

The qualitative separation of the inner region in one domain dominated by
organized orbits and another one dominated by chaotic orbits could be easily
justified by the shape of the corresponding potential well. As shown in the
left part of Fig.~\ref{fig:potential}a the potential well of the inner region in
which the orbits develop consists of a shallow part (right side of the left
well) and a much deeper part located at the innermost part of the inner region
(left side of the left well). The chaotic orbits (Fig.~\ref{FigVeffOrb}a) are
the ones that start from the shallow part of the well (see the embedded diagram
in Fig.~\ref{FigVeffOrb}b) and eventually enter the deep hole
(Fig.~\ref{FigVeffOrb}b) which works as a stochastic scatterer of the orbit. In
Fig.~\ref{FigVeffOrb}b we see only the upper part of the hole which in full
scale reaches the value of $-10^5$ for the chaotic orbit seen
in Fig.~\ref{FigVeffOrb}a. Chaos is generated by the abrupt changes of the
values of the $V_{\textrm{eff}}$ along an orbit. On the other hand the organized
orbits, which belong to the main island of the inner region,
(Fig.~\ref{FigVeffOrb}c) move on the periphery of the deep well
(Fig.~\ref{FigVeffOrb}d). Finally, the orbits which belong to the chain of
islands of the resonance $2/3$ (Fig.~\ref{FigVeffOrb}e) are orbits with well
tuned initial conditions in the phase space that start from the shallow region,
eventually enter slightly in the deep region but again move mostly on the
periphery of the hole itself (Fig.~\ref{FigVeffOrb}f), thus maintaining their
organized character.

\subsection{The neck}

\begin{figure}[htp]
 \centerline{\includegraphics[width=12pc] {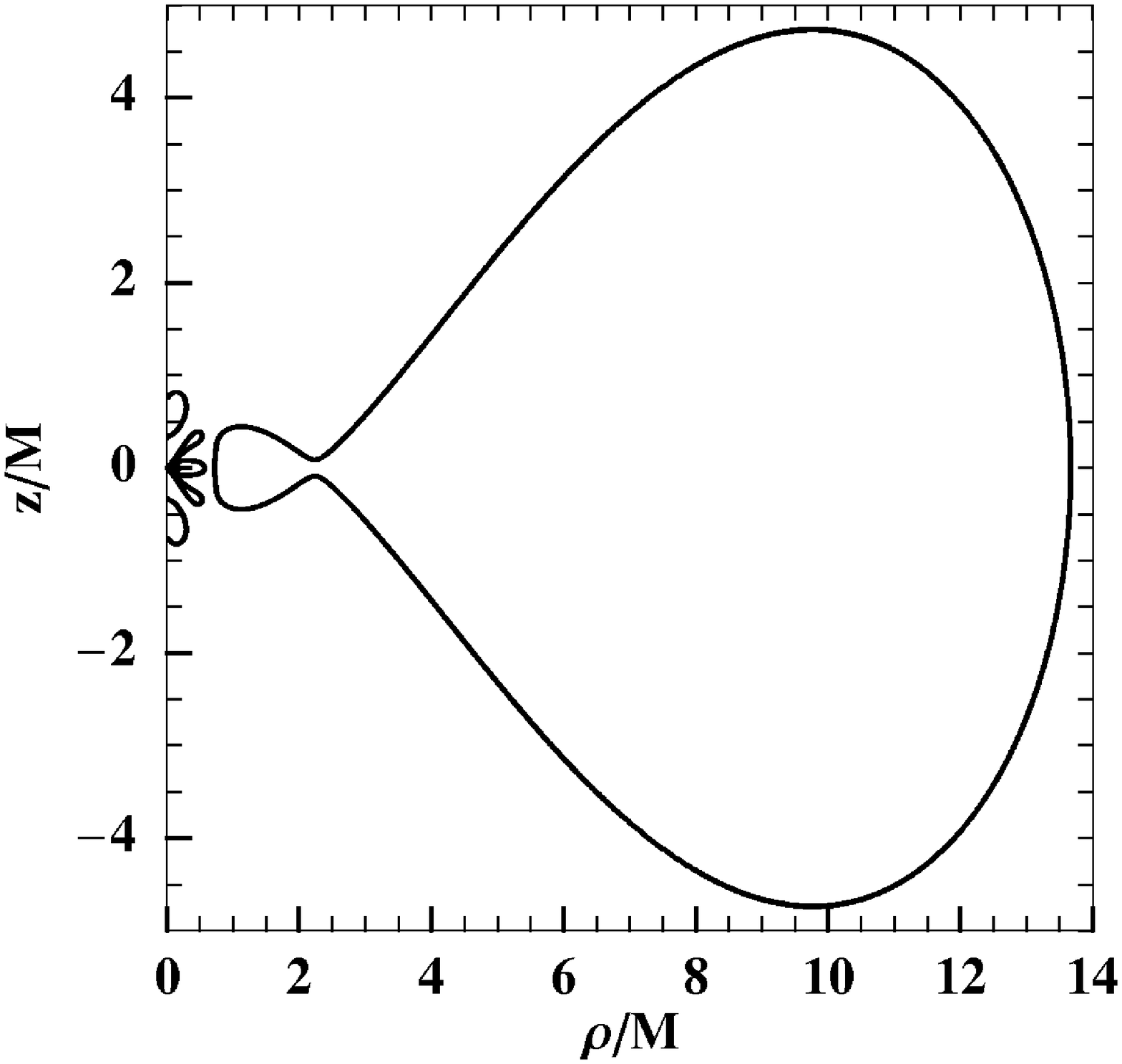}}
 \caption{The CZV for parameter values $E=0.95,~L_z=2.995 M,~\chi=0.9,~q=0.95$,
 where the outer and the inner region are connected through a short neck.
 }
 \label{FigMerGS}
\end{figure}

Previously we studied a case in which the outer and the inner regions are
separated. Let us see now what happens when the two regions communicate with
each other. In order to attain this we have slightly reduced the angular
momentum to the value $L_z=2.995 M$ without altering the other parameters. Then,
as shown in Fig.~\ref{FigMerGS} a short neck that connects the two regions
appears at $\rho\approx 2.25 M$. Through this neck, orbits are allowed to move
from the outer to the inner region and vice versa.

\begin{figure}[htp]
 \centerline{\includegraphics[width=36pc] {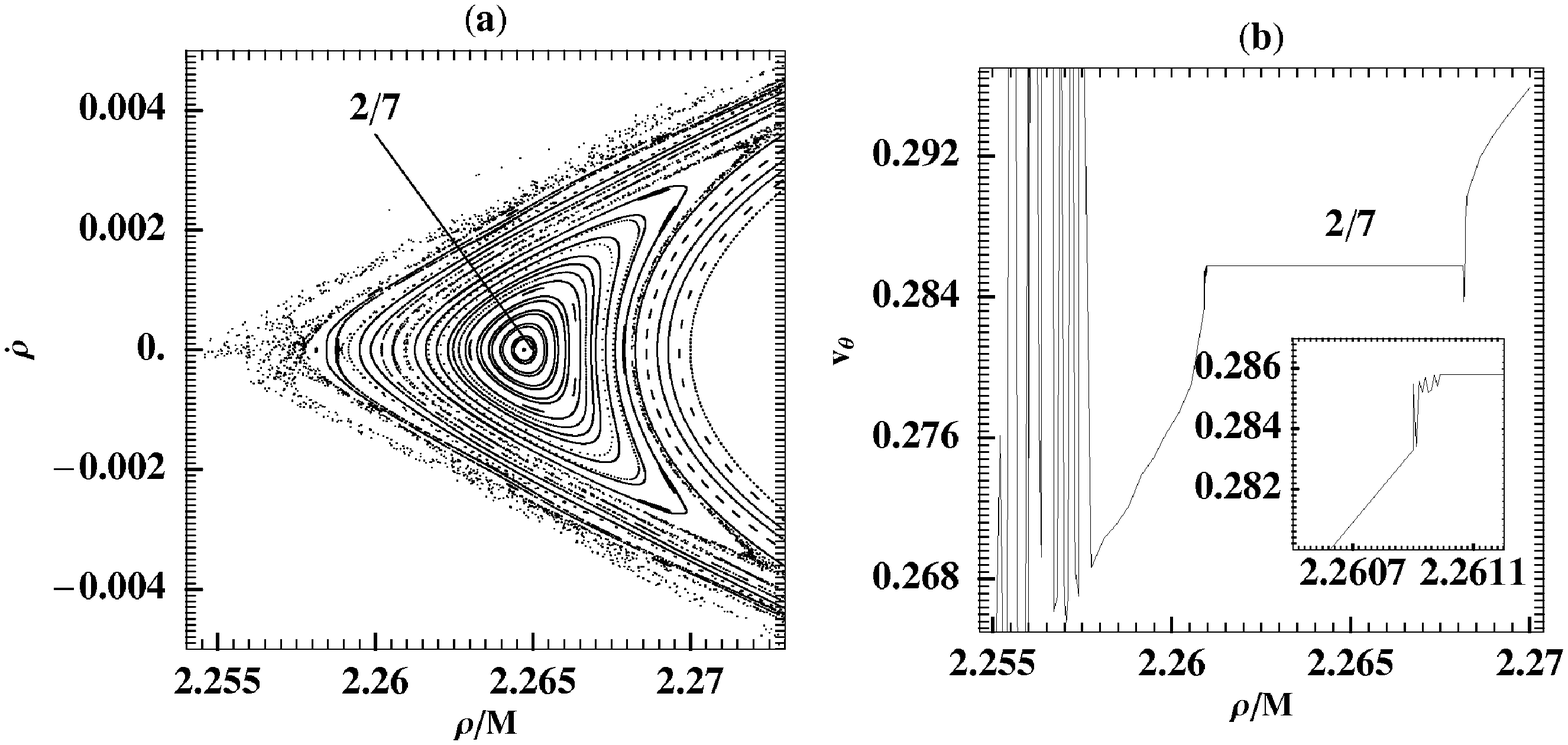}}
 \caption{ (a) A detail of the surface of section in the neighborhood of the
 neck which joins the inner and the outer regions. (b) The rotation curve along
 the $\dot{\rho}=0$ line of the surface of section presented in (a). Embedded in
 (b) is the irregular variation of the rotation number just outside the left
 side of the $2/7$-plateau. This irregular behavior is due to the chaotic layer
 surrounding the corresponding island.
 }
 \label{FigNeck}
\end{figure}

Although the main characteristics of the two regions have not changed, the
``neck'' has allowed the formation of heteroclinic sections of the asymptotic
curves emerging from unstable periodic orbits lying in the two initially
separate regions. Such sections lead to stronger chaos. In our case chaotic orbits
that were initially confined between the boundaries of the outer region and the
last KAM of the corresponding main island are able to explore the phase space of
both regions by escaping through the neck to the more chaotic inner region and
returning again through the neck to the thin chaotic layer near the periphery of
the outer region. The chaotic layer becomes more prominent when the inner and
the outer regions come closer to each other. There is also a well-known
phenomenon in nonlinear dynamics called ``stickiness'' first reported in
\cite{Contop71} (for a review see \cite{Contop02}), which could explain the
``strange'' behavior reported in \cite{Gair08}. The stickiness concerns chaotic
orbits which for various reasons stick for a long time interval in a region,
close to an invariant curve, so that their behavior on a surface of section may
resemble that of regular orbits, before extending further away.

The situation described in the previous paragraph is depicted in
Fig.~\ref{FigNeck}a which is a $z=0$ surface of section of the region of $\rho$
near the neck. The main island of the outer region is surrounded by a chaotic
layer which communicates with the chaotic sea of the inner region. In this
``outer'' chaotic layer many high-multiplicity islands of stability are present.
The borders of this chaotic layer with the region of regular orbits is quite
densely populated by chaotic orbits (Fig.~\ref{FigNeck}a). In fact the chaotic
orbits near these borders remain close to the regular orbits for a long time
before escaping to the main chaotic domain. The sticky zones are near these
borders. In order to further explore the area of the outer region near the neck
we computed the rotation number along the $\dot{\rho}=0$ line
(Fig.~\ref{FigNeck}b). On the left side of Fig.~\ref{FigNeck}b the irregular
variations of the rotation number confirm the chaoticity of the orbits that
surround the main island of the outer region. On the right of these irregular
variations the rotation curve seems to be strictly monotonic, until we reach the
resonant island $2/7$ (the plateau in the middle of Fig.~\ref{FigNeck}b).
Around the resonant island $2/7$ a thin chaotic layer can be discerned on the
surface of section, mainly around two unstable points near $\rho\simeq 2.275 M$,
$\dot{\rho}\simeq\pm 0.0035$ in Fig.~\ref{FigNeck}a. The chaotic nature of the
orbits in that layer produces again an irregular variation of the rotation
number (embedded diagram in Fig.~\ref{FigNeck}b). On the right of the $2/7$
resonant island the rotation number seems to grow like a strictly monotonic
function again.

The fact that chaotic orbits belonging to thin chaotic layers, like the ones
described above, tend to stick near regular orbits has a side-effect on their
frequency spectrum. As long as the orbits stick near a regular orbit they tend
to get two main frequencies like the frequencies corresponding to the regular
orbits. When they move away from this regular orbit they lose these two main
frequencies and the chaotic noise in their frequency spectrum prevails. However,
the orbits can approach again a regular orbit, even the one they departed from,
stick around it again for a certain interval of time and exhibit again two main
frequencies. If these regular orbits belong to resonances the ratio of the two
fundamental frequencies will be for some time a rational number, otherwise it
will be irrational. The appearance and disappearance of the two main frequencies
characterizes the existence of sticky chaotic orbits, thus it signals the
existence of non-Kerr spacetime, although this would be a difficult
observational task due to instrumental noise in gravitational wave signals.
Therefore, we mainly focus our observational method on the regular orbits which
correspond to resonances.

\section{Inspiraling orbits}
\label{sec:4}

According to general relativity an object of mass $\mu$ orbiting in the
spacetime background of a very massive object loses energy and angular momentum
by emitting gravitational waves. In section \ref{sec:3} we investigated the
characteristics of a geodesic motion in a stationary MN spacetime background.
The energy $E$ and the angular momentum $L_z$ were regarded as constants of
motion instead of adiabatically varying quantities. The accurate solution of the
two body problem in the framework of general relativity is still intractable. It
can only be solved numerically, although for a wide range of parameters the
problem has not been adequately analyzed yet. Various approximate schemes have
been proposed to compute the energy and angular momentum loss for a small mass
in a Kerr metric while no other systematic way to compute such losses for other
kinds of background spacetimes is available. Following the hybrid approximative
method of \cite{GairGlampedakis06} (Eqs.~(44,45)), that Gair et al \cite{Gair08}
also applied in the MN spacetime, we have computed the energy and $z$-angular
momentum losses from the instantaneous geometric orbital parameters as if the
spacetime were a Kerr metric. The only modification introduced by Gair et al
\cite{Gair08} from the original hybrid model \cite{GairGlampedakis06} was that
the deviation of the quadrupole moment of the MN was added to the terms that
are proportional to the square of the spin of the Kerr metric. We applied
exactly the same modification. Then we assumed linear variations for the
parameters $E$, $L_z$, that is
\be
E(t)=E(0)+\left. \frac{dE}{dt} \right|_{0} t
\ee
and
\be
L_z(t)=L_z(0)+\left. \frac{dL_z}{dt} \right|_ {0} t
\ee
and we inserted them in the equations of motion, thus producing a new set of
non-geodesic equations of motion that approximately describe the adiabatic
inspiral. To be more specific, the energy and angular momentum losses
$(dE/dt)|_0,~(dL_z/dt)|_0$ were computed from the semi-latus rectum and
eccentricity of the geodesic motion that correspond to the initial values $E(0)$
and $L_z(0)$ (see Appendix \ref{AppA}). Then the non-geodesic orbit, that
was created from the adiabatically varying equations of motion, was evolved for
an interval of time that was long compared to all orbital periods (azimuthal and
polar, e.g. $T_\rho$ and $T_z$), but short compared to the characteristic time
of the adiabatic variations $E(0)/(\frac{dE}{dt})|_0$ and
$L_z(0)/(\frac{dL_z}{dt})_0$ (the latter comparison justifies the use of linear
dependence for $E(t)$ and $L_z(t)$). In fact the evolution of the orbit was
calculated for a time interval sufficient for the orbit to hit a resonance,
cross it, and depart from it. Since the resonances we have found in exploring
the geodesics in a MN metric are very thin structures, the corresponding times
just to cross them were indeed very short in comparison to the characteristic
time of $E$-variation and $L_z$-variation. In order not to waste numerical time
and remain in the range of validity of the linear approximation for the
adiabatic changes, we chose initial conditions for the non-geodesic orbits along
the line $(\dot{\rho}=0,~z=0)$ that were within a short distance away from the
boundary of a resonant island itself.

The initial conditions along the aforementioned line give an adequately
representative set of non-geodesic orbits to study the time it takes for the
inspiraling orbits to cross through the resonant islands. Actually the
non-geodesic orbits corresponding to the initial conditions that lie on the
convex side of the islands (Fig.~\ref{FigExtReg}a) do not cross the particular
resonant island; these orbits continuously recede from the islands. Thus from
now on we constrain our study to non-geodesic orbits that start near the concave
side of the resonant islands. On the concave side of the leftmost island there
is a window of initial conditions of $\rho(0)$ along the $(\dot{\rho}=0,~z=0)$
line for which the corresponding non-geodesic orbits eventually enter first this
particular island of resonance. Upon its entrance a non-geodesic orbit visits
all the other islands of that resonance as well. An analogous window of
$\rho(0)$s, right next to the former one, contains orbits that enter first the
next island of the same resonance and so on until we reach the window that
enters first the last island of the chain. After that window, it follows another
window of initial conditions for which the non-geodesic orbits enter the
leftmost island first. The non-geodesic orbits that start from the latter window
perform an extra full rotation on the surface of section with respect to the
non-geodesic orbits of the first window, before they enter the  particular
resonant island. During this rotation the islands move towards the center of the
main island, so that they eventually encounter the evolving orbits from this new
window. The succession of such windows continues, until point that the
non-geodesic orbits reach the resonance is so close to the center of the main
island (due the energy and $z$-angular momentum loss), that the islands have
shrunk to negligible thickness and thus no resonance can be observed. However
the linear approximation we use to produce the non-geodesic orbits ceases to be
valid well before this happens. Therefore we have not extended our numerical
computation that far.

\begin{figure}[htp]
 \centerline{\includegraphics[width=20pc]
{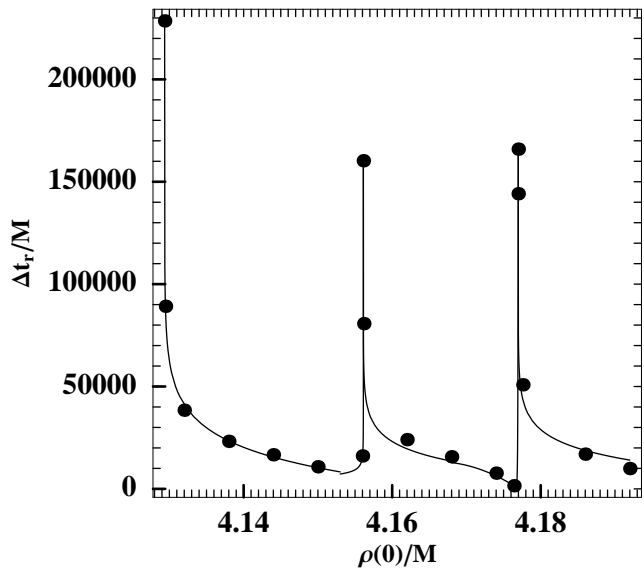}}
 \caption{ The time $\Delta t_r$ needed by non-geodesic orbits to cross the
 chain of islands belonging to the $2/3$-resonance as a function of their
 initial conditions $\rho(0)$ (the initial value of the $\rho$-coordinate) along
 the line $\dot{\rho}=0, z=0$. The parameters used are
 $\mu/M=8 \times 10^{-5},~q=0.95, ~\chi=0.9,~E(0)=0.95,~L_z(0)=3 M$.
 }
 \label{FigTime}
\end{figure}

In order to study the observability of the resonances we have chosen a few
distinct initial conditions along the above sequence and measured the time
interval that the corresponding non-geodesic orbit needs to cross the
$2/3$-resonant islands. The time $\Delta t_r$ spent by each non-geodesic orbits
in the $2/3$-resonance is shown in Fig.~\ref{FigTime} as a function of their
initial conditions along the line $(\dot{\rho}=0,~z=0)$. Moreover each window in
the sequence of windows in Fig.~\ref{FigTime} is demarcated by two distinct
abrupt peaks of $\Delta t_r$ lying on either side of the window. Of course there
are a number of parameters that affect this $\Delta t_r$ time interval, like the
initial energy $E(0)$, the initial angular momentum $L_z(0)$ and the exact
initial conditions of an orbit (in our case $\rho(0)$), as well as the physical
characteristics of the EMRI, namely the parameters of the metric $\chi$ and $q$,
and the masses of the two bodies. Especially the ratio of masses $\mu/M$ plays
an important role in $\Delta t_r$; the lower the value of $\mu/M$ is, the longer
time it takes to reach a particular island, and consequently the width of the
windows is larger. Our aim is not to explore the dependence of crossing time
$\Delta t_r$ on all these parameters, instead we explore the qualitative
characteristics that relate the time of crossing with the physical
characteristics of an EMRI. We observed that the exact point of entrance in the
island is crucial for the time the orbit spends in that island before it exits
the particular chain of islands.

\begin{figure}[htp]
 \centerline{\includegraphics[width=36pc] {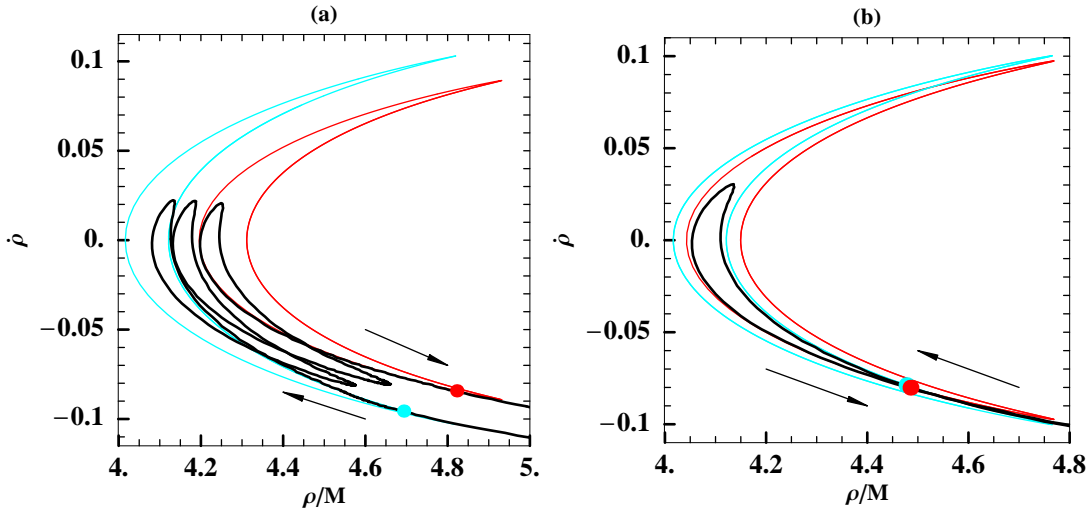}}
 \caption{The black thick lines mark the succession of every third point
 on the $z=0$ surface of section of a non-geodesic orbit for two different types
 of entrapment in the $2/3$-resonance. (a) Entrapment of a non-geodesic orbit
 that makes 3 loops while inside the resonant island and (b) entrapment of a
 non-geodesic orbit that makes almost one loop. The arrows show the flow
 of the orbits on the $\rho,\dot{\rho}$ plane. In (a) and (b) the blue (light
 gray) big dot indicates the point at which the corresponding non-geodesic
 orbit crosses the border (blue or light gray thin line) and enters the leftmost
 island of the $2/3$-resonance, while the red (dark gray) big dot indicates the
 point at which the non-geodesic orbit crosses the border (red or dark gray thin
 line) and exits the $2/3$-island of stability. Note the drift of the island
 during the evolution of the orbit. For both cases
 $\mu/M=8 \times 10^{-5},~q=0.95,~\chi=0.9,~ E(0)=0.95,~L_z(0)=3 M$.
 The non-geodesic orbits of (a) and (b) correspond to the first and the third
 point from the left on the diagram of Fig. \ref{FigTime} respectively.
 }
 \label{FigOrCro}
\end{figure}

In order to study how the time spent in the chain of islands depends on the
entrance point we use a stroboscopic depiction, i.e. for a chain of islands of
multiplicity 3 we depict only every third crossing of an orbit through a surface
of section. Consequently the successive crossing points are close to each other
and we can join them by a single line (thick line in Fig.~\ref{FigOrCro}a,b).
In this figure only one island is depicted; the leftmost one of the
$2/3$-resonance (Fig.~\ref{FigExtReg}a). The stroboscopic projection of a
non-geodesic orbit forms a clockwise spiral on a surface of section until it
reaches the resonant island. By plotting a couple of such orbits, namely one
entering the island very close to its lowest point (Fig.~\ref{FigOrCro}a) and a
second one at a point a little higher from its lowest point and along its
concave side (Fig.~\ref{FigOrCro}b), we noticed that if the entrance point is
near the lowest point of the island it will spend substantially more time in the
resonance than when the entrance point is further upwards. In the former case
(Fig.~\ref{FigOrCro}a) the adiabatic drift of the orbit and of the island itself
forces the orbit to revolve counterclockwise three times around the island while
at resonance. After these revolutions the relative drift of the orbit with
respect to the island moves the orbit out of the island and the exit point is
then on the right of the entrance point. In the latter case
(Fig.~\ref{FigOrCro}b) while the orbit starts revolving counterclockwise around
the island upon its entrance, it finds itself near the opposite side (the convex
side) of the island when the relative drift forces the orbit to exit the island.
In this case the exit point is on the left of the entrance point but very close
to it.

As mentioned before, the ratio of masses that are involved plays a radical role
in the crossing time. If this ratio is  $\gtrsim 10^{-4}$ the evolution of the
orbit is so quick that within one $z$-period the orbit has crossed the resonance
and almost no point on a surface of section is found inside the island. On the
other hand if $\mu/M \lesssim 10^{-7}$ the adiabatic evolution of the orbit is
so slow that the determination of the initial conditions that eventually enter
an island of resonance is quite difficult due to long integration times.
Interestingly enough, the range of mass ratios that lead to observable crossing
times correspond to the range of masses of EMRIs for which LISA is expected to
be sensitive.

The next important question one has to answer is how probable is for a non-Kerr
EMRI to evolve so as to pass through a resonance during its evolution, and thus
enable us to detect such a passage through a frequency analysis of the
corresponding gravitational waves. Using again the MN metric as a characteristic
example of a  perturbed Kerr metric, we have explored how the initial conditions
evolve if the orbit starts far away from a specific resonance. We have not
explored all possible orbital parameters to check if they finally pass through
such a resonance, since the problem depends on too many parameters: the ratio of
masses $\mu/M$, the quadrupole deviation parameter $q$, the spin $\chi$, the
initial orbital characteristics (the semi-latus rectum $p$, the eccentricity $e$
and the inclination of the orbit $\iota$) and the particular resonance we
examine each time. Instead of calculating the evolution of the non-geodesic
orbits for all these parameters we  run an orbit with a specific initial
condition to obtain a crude estimate of what are the chances to hit a resonance.
More specifically we focused our attention to the resonance 2/3 which according
to our previous analysis has the strongest effect (the corresponding islands are
the thickest ones). We started with initial values
$E(0)=0.95,~L_z(0)=3 M,~q=0.95,~\chi=0.9$. For these values there are two
allowed regions of orbits, while the outer region is characterized by an
effective potential that is qualitatively similar to the one for a Kerr metric
with the same mass and angular momentum. The leftmost island of the
$2/3$-resonance (see Fig.~\ref{FigExtReg}a) of such a MN spacetime then spans
the interval $4.006 M \leq \rho_{2/3} \leq 4.109 M$. 

We chose an initial orbit with the above parameters and evolved it by varying
adiabatically the parameters $E,~L_z$. The rates of their variation was computed
based on the initial orbital parameters and the corresponding Kerr-like losses
(see the discussion above). The surface of section of the non-geodesic orbit was
drawn and the speed of the drift of the corresponding KAM curves was graphically
calculated; i.e., we measured the velocity by which the leftmost part of the
orbit on a surface of section corresponding to a non-geodesic orbit was moving.
Due to energy and $z$-angular momentum loss the corresponding geodesic KAM
curves are shrinking (circularization of the orbits). Simultaneously they are
moving towards lower $\rho$-values. The net drift of the leftmost part of these
geodesic curves is towards $\rho=0$ with a speed of the order of
$\Delta \rho_{orb}/\Delta t \simeq -1.6 \times 10^{-6}$ (the minus sign means
that it moves towards $\rho=0$). This number corresponds to a ratio of masses
$\mu/M=8 \times 10^{-5}$ and initial values for the orbit
$\rho(0)=4.15~M,\dot{\rho}(0)=0,z(0)=0$. Of course this drift is expected to be
slower for lower $\mu/M$ values and to be slightly altered for other values
of $\rho(0)$ due to changes in the corresponding orbital parameters $p,e,\iota$
that consequently induce changes in the values of $E$ and $L_z$ losses. On the
other hand we followed the successive positions of the chain of islands for the
resonance $2/3$ due to the values of $dE/dt,~dL_z/dt$ used in the previous
calculation. The whole chain is shrinking and approaches the center of the main
island. But the center of the main island itself is moving, together with the
set of islands around it, towards $\rho=0$. The net drift of the leftmost island
is away from $\rho=0$ with velocity of order
$\Delta \rho_{isl}/\Delta t \simeq +1.1\times 10^{-6}$. Therefore the orbit will
hit the 2/3-resonance in a time interval of order $4 \times 10^5 \Delta\rho_0$,
where $\Delta\rho_0$ is the initial distance between the  value of $\rho$
for the orbit when it crosses the $z=0,\dot{\rho}=0$ surface, and the right
boundary along the axis $\dot{\rho}=0$ of the leftmost $2/3$-resonant island.

This crude estimate for the time to hit the 2/3 resonance has its own range of
validity. While the orbit and the chain of islands shrink towards the center of
the main island, the resonance is eventually led to disappearance. As the chain
of islands approaches the central point of the main island, the islands of the
$2/3$-resonance shrink and disappear at the center of the main island. Therefore
we should not extend the estimate for the time to hit the resonance beyond the
time when there are no 2/3-resonance islands at all. This upper value for the
time to hit the resonance is directly related to the lowest eccentricity of
orbits that will eventually hit the resonance (the closer the orbits are to the
central point of the main island, the smaller is their oscillation along the
$\rho$ axis). If the orbit starts with eccentricity lower than a cutoff value,
the orbit will become very circular before it reaches a ratio of frequencies
corresponding to the resonance of 2/3. All higher initial eccentricities, up to
the eccentricity that corresponds to the initial position of the chain of
islands itself will eventually cross this chain of islands, leaving a
characteristic imprint on the ratio of gravitational wave frequencies. For the
aforementioned initial values the least eccentric orbit that has sufficient time
to hit the $2/3$-resonance has initial conditions
$\rho(0)\simeq 6.8~M,\dot{\rho}(0)=0,z(0)=0$ (the rotation number of
the corresponding geodesic orbit is $\nu_\theta(0)=0.26$). This initial
condition corresponds to orbital parameters
 $p(0)=9.1~M,~e(0)=0.16,~\iota(0)=27^\circ$. The time needed by such an orbit
to hit the $2/3$-resonance is of order $5 \times 10^5~M$, that is
$\simeq 2.5 (M/M_\odot) ~\textrm{s}$. On the other hand the most eccentric orbit
that hits the $2/3$-resonance in a very short time interval (since it starts
right next to the concave side of the island) has orbital parameters
$p(0)=7.4,~e(0)=0.46,~\iota(0)=19^\circ$ ($\nu_\theta(0)=0.33$).

Although this single numerical example represents an arbitrary initial orbital
configuration, it still gives a crude estimate of the orbital characteristics
that have the chance to reveal a possible non-Kerr character of the background.
We expect that the value of $\mu/M$ will not alter significantly the range of
orbital parameters since it will simply make the evolution faster (large
$\mu/M$) or slower (low $\mu/M$); it will only affect the total time to hit the
resonance. The initial parameters $E(0), L_z(0)$ will be more crucial for
determining the range of parameters since they are directly related to the
initial position of the islands of resonance.

\section{Observational imprints}
\label{sec:5}

In this section we focus our attention on the quantity that, according to our
analysis, its measurement will provide us with useful information about possible
deviations of an EMRI's central object from a Kerr black hole. This quantity is
the ratio of the polar fundamental frequencies, that is the frequencies that
characterize the oscillations of an orbit on the polar plane which rotates along
with the low-mass object.

 \begin{figure}[htp]
 \centerline{\includegraphics[width=32pc] {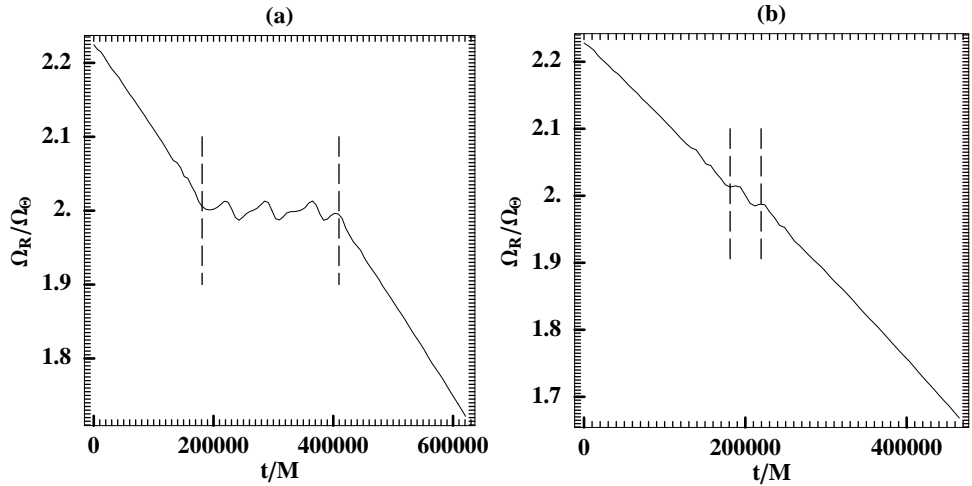} }
 \caption{ The evolution of the ratio $\Omega_R/\Omega_\Theta$ as function of
 the coordinate time $t$ for (a) the non-geodesic orbit shown in
 Fig.~\ref{FigOrCro}a and (b) the non-geodesic orbit shown in
 Fig.~\ref{FigOrCro}b. The vertical dashed lines  demarcate the time intervals
 that the non-geodesic orbit spends in the interior of the $2/3$-resonance.
 }
 \label{FigRevol}
\end{figure}

For a generic EMRI signal each fundamental frequency related to the evolution of
the system, as well as any linear combination of their harmonics, will show
up in its Fourier analysis. The most prominent frequency peaks $\Omega_k$ in the
spectrum will be some integer multiples of the corresponding fundamental
frequencies $\omega_i$; that is
$\Omega_k=\displaystyle\sum_i m^{(k)}_i \omega_i$,
where $m^{(k)}_i$ are integers. Therefore the ratio of the frequencies
$\Omega_k$ of the most intense peaks that are observed in the Fourier spectrum
of the signal and are related to polar oscillations, but which are not
harmonically related to each other, will be of the form
\be
\frac{\Omega_\rho}{\Omega_z}
=\frac{m_1 \nu_\theta + m_2}{n_1 \nu_\theta + n_2},
\ee
with $m_1,m_2,n_1,n_2$ some integers. In our case the highest peaks in the
Fourier spectrum of the two polar coordinates yield a ratio of Fourier
frequencies equal to $(\nu_\theta-1)/\nu_\theta$
($m_1=1,m_2=-1,n_1=1,n_2=0$).

The Fourier spectrum of the coordinates used to describe the polar oscillations
consists of both fundamental frequencies and combinations of their harmonics, as
mentioned before. This may render the determination of Fourier frequencies
problematic since we practically analyze a finite length of an orbit that leads
to Fourier peaks of finite width; thus two frequency peaks that are close to
each other may not be easily discerned. In order to overcome this problem, we
used the coordinates $R$ (see Eq.~(\ref{fR})~) and
\be
\Theta=\tan^{-1}\left( \frac{z}{\rho} \right)
\ee
instead of the $\rho,z$ coordinates. As shown in Appendix \ref{AppB} the new
coordinates are much better in order to reveal their frequency content with good
accuracy. Thus the frequencies used in the following paragraphs have been
computed by Fourier analyzing the time series of the orbits described by these
''cleaner'' coordinates $R$ and $\Theta$.

As explained in Sec.~\ref{sec:3.1}, if an EMRI source evolves in the
gravitational field of a massive object that is similar to, but not exactly a
Kerr black hole, it has good chances to pass through a discernible resonance
 --like the $2/3$-resonance, for which some quantitative estimates have been
given in the previous section. If this happens within the range of distances
that a gravitational-wave detector like LISA is sensitive to detect a
corresponding signal \cite{LISA}, we should be able to observe the non-Kerr
character of the central object by tracking a transient stationary value of the
ratio of frequencies $\nu_\theta$.

We have evolved a few orbits that initially lie near a resonance. We have
avoided to investigate orbits that start far away from a resonance since this
would be quite expensive numerically. Also the time until the inspiraling orbit
hits a resonance is of no special interest since the evolution of the orbits
outside a resonance is expected to be quite similar both in a Kerr and in a
perturbed Kerr metric; in both cases the ratio of frequencies will vary
monotonically with time.

As explained in Sec.~\ref{sec:4} the passage through a resonance depends
crucially on the location of the entrance point in an island of resonance.
Therefore we will present two graphs that depict the evolution of the ratio of
frequencies as a function of time. The first one corresponds to an entrance
point, such that the orbit gets trapped for a few circles in the islands of the
2/3-resonance. The second one corresponds to an entrance point, such that the
orbit performs almost one loop and then departs from the chain of islands. The
two cases are exactly the ones that were presented in Sec.~\ref{sec:4} when we
discussed the duration of crossing a resonance as a function of the initial
conditions (cf.~Fig.~\ref{FigOrCro}a and b). In both cases the orbits have been
evolved for sufficiently long time before and after their entrance, so that the
different type of evolution of the ratio of frequencies then is clear.

\begin{figure}[htp]
 \centerline{\includegraphics[width=15pc] {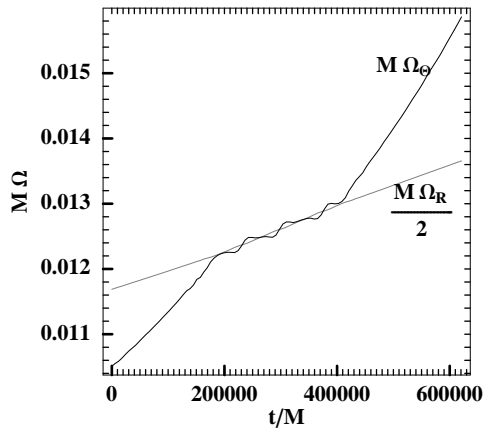} }
 \caption{ The evolution of the two polar frequencies $\Omega_{R}$ and
 $\Omega_\Theta$ as functions of the coordinate time $t$ for the non-geodesic
 orbit shown in Fig.~\ref{FigOrCro}a before, during and after the crossing of
 the $2/3$-resonance
 }
 \label{FigFre}
\end{figure}

In Fig.~\ref{FigRevol} we have plotted the ratio of frequencies (i) for an orbit
that its surface of section evolves through an island of 2/3-resonance according
to Fig.~\ref{FigOrCro}a and (ii) for an orbit that corresponds to
Fig.~\ref{FigOrCro}b. The orbits have been divided in time segments of length
$\Delta t \simeq 5000 ~M$. Each such segment of the orbit is Fourier analyzed
and the frequencies $\Omega_R$, $\Omega_\Theta$ are recorded (see discussion
above). Finally the ratio of these frequencies is plotted as a function of $t$.
For the case in Fig.~\ref{FigOrCro}a we have plotted also the evolution of the
two frequencies (Fig.~\ref{FigFre}). In this figure we have highlighted the
crossing of the $2/3$-resonance by the non-geodesic orbit by
scaling accordingly the frequency $\Omega_R$ so that the two frequencies
coincide when they are superimposed. As shown in Figs.~\ref{FigRevol} the ratio
of frequencies evolves quite differently while the non-geodesic orbit passes
through a resonance compared to its behavior when this orbit moves adiabatically
from one KAM curve to another (outside the resonance). However the ratio of
frequencies does not have the form of a pure plateau when at resonance. Instead
the ratio of frequencies forms a number of oscillations (the number of cycles
matches the number of loops that the orbit performs while it crosses a
particular island) superimposed on a plateau. These oscillations are mainly due
to the finiteness of the orbit that is Fourier analyzed. In most of the cases
the orbit performs only part of a loop inside an island though
(Fig.~\ref{FigOrCro}b). Then there is only a fraction of an oscillation apparent
on the evolution of $\Omega_R/\Omega_\Theta$ and the variation of the curve may
not show very clearly the underlying plateau (Fig.~\ref{FigRevol}b). A possible
way to discern the plateau in these cases is by extrapolating the lines that
describe the variation of $\Omega_R/\Omega_\Theta$ before and after the plateau
and look for some finite distance between these lines. This distance will be a
measure of the time spent within a resonance and should be related to the
characteristics of the source.

\section{Conclusions} \label{sec:6}

In this paper we have investigated what are the manifestations of EMRIs
consisting of a small mass compact object that is orbiting around a much more
massive compact object that is not an exact Kerr black hole. Our study has been
based on a generic qualitatively new feature that discerns an integrable
Hamiltonian system from a system that slightly deviates from an integrable one;
namely, the appearance of Birkhoff chains of islands on a Poincar\'{e} surface
of section, instead of a set of fixed points of the corresponding integrable
system. We have chosen a specific exact solution of the vacuum Einstein
equations, a member of the family of the Manko-Novikov  metrics, which has been
extensively studied in the literature. This MN metric is characterized, besides
its mass $M$ and spin $S$, by one more parameter $q$, which measures the
deviation of its quadrupole moment from the corresponding Kerr (the one with the
same mass $M$ and spin $S$). Thus if the $q$ parameter is set to zero the
Manko-Novikov metric turns into an exact Kerr metric. We have used this metric
as a generic example of a slightly non-Kerr metric.

By studying first the geodesic orbits in a MN background, we showed that when
the orbit hits a resonance, that is when the ratio of the frequencies
$\omega_\rho$, $\omega_z$ that characterize the orbital oscillations on the
polar plane (the plane that passes through the axis of symmetry of the central
object and rotates along with the low mass object) is a rational number, then
the orbit exhibits a qualitatively new behavior: for a finite range of initial
conditions the ratio of these frequencies is constant, in contrast to what
happens in the integrable case of a Kerr metric. Actually we have found two
such chains of Birkhoff islands in the most interesting region of bound orbits
(the outer one), which correspond to the resonances 2:3 and 1:2 respectively.
Furthermore we explored the thickness and the location of the $2/3$ islands with
respect to the various parameters that characterize the metric and the
parameters describing the orbits. Thus, in a realistic EMRI case with a non-Kerr
central object, while the orbit of the low-mass object evolves adiabatically,
the ratio between the corresponding frequencies is expected to remain constant
(the two frequencies get locked to each other) for a finite time interval,
exhibiting a characteristic plateau in the evolution of this ratio.

Although there is a wide range of time intervals corresponding to such a
plateau, depending on the exact parameters of the metric, the ratio of masses,
and the specific parameters and initial value of the coordinates of the orbit,
the appearance of a plateau in the evolution of the ratio of frequencies with
time is a generic feature. Moreover the most prominent plateaus correspond to
fractions of small integer numbers, since these are related to strong resonances,
and thus they are easier to investigate through focused data analysis of the
gravitational-wave signals from EMRI sources. We found that the duration of the
plateau through the resonance of 2:3 is roughly of the order of
$\Delta t_r \approx 0.15 (M/M_\odot) \textrm{s}$ for a ratio of masses
$\mu/M=8 \times 10^{-5}$ and a MN metric with $q=0.95$ and $\chi=0.9$. This time
$\Delta t_r$ increases for lower ratios of masses and decreases for higher
ratios of masses. For $\mu/M \gtrsim 10^{-4}$ the evolution of the orbit through
a resonance is so quick that the corresponding plateau is rather impossible to
be detected though.

The initial orbital parameters (semi-latus recta $p$, eccentricities $e$, and
inclinations $\iota$) that will eventually lead the orbit through a strong
resonance apparently span quite a wide range; hence a high fraction of possible
sources for LISA are expected to exhibit a plateau if they involve such a
non-Kerr central object. A rough estimate yields $e$'s, and $\iota$'s in the
range of $0.17-0.45$ and $44^\circ - 19^\circ$ respectively, corresponding to
initial semi-latus recta of the order $7.4 M - 9.1 M$, that will evolve so as to
hit the resonance 2:3.

We believe that a focused data analysis of signals of LISA in  the temporal
region where the fundamental frequencies related to the orbital oscillations on
the polar plane of the EMRI's orbit has high chances to reveal or at least
constrain the non-Kerrness of the EMRI central object. This analysis, along with
a number of other tests that have been proposed by other people
\cite{CollHugh04,GlamBaba06,Gair08}, and are related with the non-Kerrness of
the metric involved, could enhance our knowledge about the astrophysical
processes that lead to creation of ultra-compact supermassive objects at the
centers of galaxies.

\begin{acknowledgments}
We would like to thank Mr. John Deligiannis for an independent cross-checking of
our first numerical results. G.~Lukes-Gerakopoulos was supported in part by the
I.K.Y. scholarships and by the Research Committee of the Academy of Athens.
T.~Apostolatos would like to thank  K.S. Thorne and I. Mandel for very
enlightening discussions on the subject. Also T.~Apostolatos acknowledges
the research funding program ``Kapodistrias'' of ELKE (Grant No 70/4/7672)
and the I.K.Y.~(IKYDA 2010) 
\end{acknowledgments}

\appendix

\section{Computation of the orbital parameters}
\label{AppA}

In order to compute the energy and $z$-angular momentum losses of the EMRI due
to gravitational radiation we have used the hybrid model of Gair and Glampedakis
\cite{GairGlampedakis06} which have actually been constructed to measure the
corresponding fluxes at infinity of an EMRI in a Kerr background. These
instantaneous fluxes, described by formulae (44,45) of their paper, are given as
functions of the orbital parameters $p,e,\iota$ of the corresponding geodesic
orbit of a test particle in a Kerr metric. When the geodesic orbit is described
in Boyer-Lindquist coordinates the aforementioned orbital parameters are given
by analogy with the Keplerian quantities:
\begin{eqnarray}
p=\frac{2 r_+ r_-}{r_+ + r_-},  \label{SemLatRec} \\
e=\frac{r_+ -r_-}{r_+ + r_-}, \label{eccentr} \\
\iota=\frac{\theta_+ -\theta_-}{2}, \label{inclin}
\end{eqnarray}
where $r_+~(r_-)$ are the maximum (minimum) radial values, while
$\theta_+~(\theta_-)$ are  the maximum (minimum) values of $\theta$ coordinate
along the orbit.

The MN metric, and consequently the geodesic orbits that we computed numerically,
are expressed in cylindrical coordinates. Thus in order to estimate the orbital
parameters, we have transformed the $\rho,z$ coordinates in Boyer-Lindquist
coordinates (the coordinates in which the Kerr metric that we obtain when we set
$q=0$ in our MN metric yields its usual form \cite{MTW}) according to
\begin{equation} \label{BoyLind}
r/M=1+\sqrt{-\sigma_2+\sqrt{\sigma_2^{~2}+\sigma_1*(z/M)^2}},~~~
\theta=\cos^{-1} \left( {\frac{z/M}{(r/M)-1}} \right),
\end{equation}
where
\begin{equation*}
\sigma_1=\chi^2-1,~~~\sigma_2=\frac{\sigma_1-(\rho/M)^2-(z/M)^2}{2}.
\end{equation*}
%

\section{Better coordinates for Fourier analysis}
\label{AppB}

The gravitational waves from an EMRI source are expected to have the same
Fourier spectrum as the orbit itself. Thus by monitoring a gravitational wave
from such a source we can reveal all the frequencies that are related with the
orbit. By using a specific set of coordinates to describe the orbit we may face
the following problem: Each coordinate most probably will describe a synthesis
of all oscillations. This will cause problems in the Fourier analysis of the
coordinate itself, because the sample that is analyzed has finite length as it
is the case when the adiabatically inspiraling orbit in a MN is Fourier analyzed
(see Sec.~\ref{sec:5}).

\begin{figure}[htp]
 \centerline{\includegraphics[width=36pc] {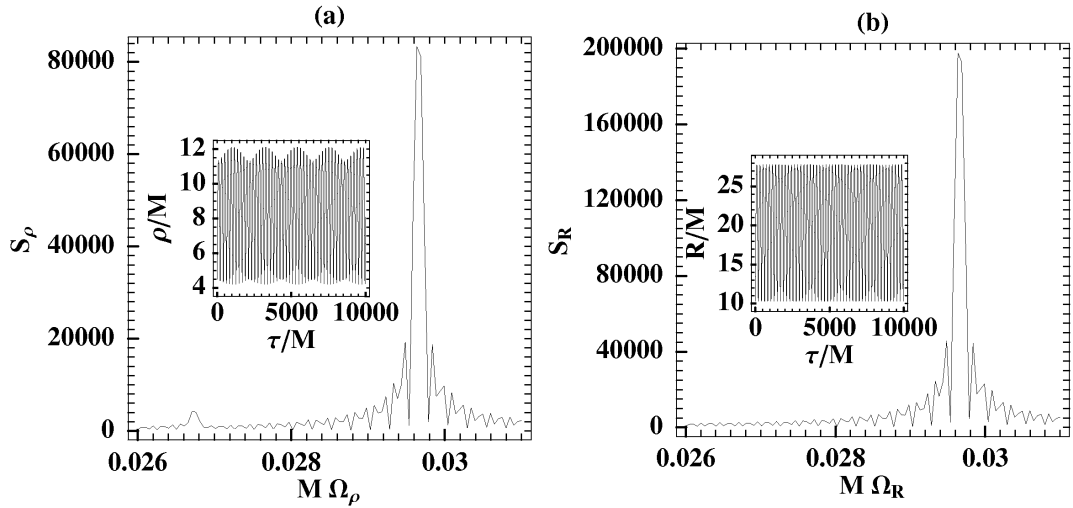} }
 \caption{ (a) The Fourier spectrum of the $\rho$-coordinate for a geodesic
 orbit. Embedded in (a) is the evolution of the $\rho$-coordinate along the
 proper time that produces this Fourier spectrum. The side frequency is apparent
 as a small secondary peak on the left of the fundamental frequency peak. The
 data-series are quite long here; namely
 $\Delta \tau_\textrm{tot}=5 \times 10^4 M$. However small the secondary peak
 may be, it causes significant problems when one tries to determine the
 fundamental frequencies of the adiabatically varying orbit by analyzing a
 data-series 10 times shorter in order to monitor the adiabatic evolution of the
 frequencies. In that case the fundamental-frequency peak is 10 times broader
 and the secondary peak slightly alters its shape. (b) The Fourier spectrum of
 the $R$-coordinate for the same geodesic orbit as in (a). Embedded in (b) is
 the evolution of the $R$-coordinate along the proper time that produces this
 Fourier spectrum. The $R$-coordinate has practically no modulation and the
 corresponding Fourier spectrum has no secondary peaks.
 }
 \label{fig:modulation}
\end{figure}

When the cylindrical coordinates are used to describe the orbital
oscillations on the polar plane there is a large uncertainty in determining the
fundamental Fourier frequencies of the corresponding coordinates. This is due to
the fact that these coordinates exhibit significant amplitude modulation. This
modulation causes the appearance of side-frequencies which may render
problematic the determination of the fundamental frequencies of a finite-length
data series (see Fig.~\ref{fig:modulation}). We found that the analysis is much
clearer in a new set of coordinates:
\begin{equation*}
R=\frac{\sqrt{\rho^2+z^2}}{k}~,~\Theta=tan^{-1} \left( \frac{z}{\rho} \right).
\end{equation*}
Thus we transformed the $\rho,z$ coordinate into $R,\Theta$ coordinates before
we Fourier analyze them, and we got a much clearer picture of the evolution of
the ratio of frequencies in the adiabatically changing orbits.

\end{document}